\documentclass[]{aastex63}

\newcommand{\NH}{N{\rm (H)}}

\received{TBD}
\revised{TBD}
\accepted{TBD}
\submitjournal{ApJ}

\bibpunct[; ]{(}{)}{,}{a}{}{;}

\def\simgt{\lower.5ex\hbox{$\; \buildrel > \over \sim \;$}}
\def\simlt{\lower.5ex\hbox{$\; \buildrel < \over \sim \;$}}

\def\s{\phantom}

\def\amin{\ifmmode^{\prime}\else$^{\prime}$\fi}
\def\asec{\ifmmode^{\prime\prime}\else$^{\prime\prime}$\fi}

\def\simlt{\mathrel{\hbox{\rlap{\hbox{\lower4pt\hbox{$\sim$}}}\hbox{$<$}}}}
\def\simgt{\mathrel{\hbox{\rlap{\hbox{\lower4pt\hbox{$\sim$}}}\hbox{$>$}}}}

\def\xmm{{\it XMM-Newton}}

\def\cygnusx1{Cygnus~X-1}
\def\cygx1{Cyg~X-1}
\def\gx339{GX~339-4}

\def\nh{{$N$(H)}}

\def\ismabs{{\tt ISMabs}}

\shorttitle{High Resolution X-Ray Spectroscopy of Fe L Absorption}
\shortauthors{Corrales et al.}
\graphicspath{{./}{figures/}}

\begin{document}

\title{High-Resolution X-Ray Spectroscopy of Interstellar Iron Toward Cygnus X-1 and GX 339-4}

\correspondingauthor{Lia Corrales, Eric V. Gotthelf, Daniel Wolf Savin}
\email{liac@umich.edu, eric@astro.columbia.edu, dws26@columbia.edu}

\author[0000-0002-5466-3817]{L\'{\i}a Corrales}
\affiliation{Astronomy Department, University of Michigan, Ann Arbor, MI 48109, USA}

\author[0000-0003-3847-3957]{Eric V. Gotthelf}
\affiliation{Columbia Astrophysics Laboratory, Columbia University, 550 West 120th Street, New York, NY 10027-6601, USA}

\author[0000-0002-3252-9633]{Efrain Gatuzz}
\affiliation{Max-Planck-Institut f\"ur extraterrestrische Physik, Giessenbachstrasse 1, D-85748 Garching, Germany}

\author[0000-0002-5779-6906]{Timothy R. Kallman}
\affiliation{NASA/GSFC, Code 662, Greenbelt MD 20771, USA}

\author[0000-0002-7336-3588]{Julia C. Lee}
\affiliation{Harvard University, John A.\ Paulson School of Engineering \& Applied Science, 29 Oxford Street, Cambridge, MA 02138, USA}

\author[0000-0002-1228-5029]{Michael Martins}
\affiliation{Institut f\"ur Experimentalphysik, Universit\"at Hamburg, Luruper Chaussee 149, 22761 Hamburg, Germany}
\affiliation{Center for Free-Electron Laser Science, Notkestrasse 85, 22607 Hamburg, Germany}

\author[0000-0003-3847-3957]{Frits Paerels}
\affiliation{Columbia Astrophysics Laboratory and Department of Astronomy, Columbia University, 550 West 120th St., New York, NY 10027, USA}

\author[0000-0002-1049-3182]{Ioanna Psaradaki}
\affiliation{Astronomy Department, University of Michigan, Ann Arbor, MI 48109, USA}

\author[0000-0002-6166-7138]{Stefan Schippers}
\affiliation{I. Physikalisches Institut, Justus-Liebig-Universit\"at Giessen, Heinrich-Buff-Ring 16, 35392 Giessen, Germany}

\author[0000-0002-1111-6610]{Daniel Wolf Savin}
\affiliation{Columbia Astrophysics Laboratory, Columbia University, 550 West 120th Street, New York, NY 10027-6601, USA}

\begin{abstract}

We present a high-resolution spectral study of Fe $L$-shell extinction by the diffuse interstellar medium (ISM) in the direction of the X-ray binaries \cygnusx1\ and GX~339--4, using the \xmm\ reflection grating spectrometer. 
The majority of interstellar Fe is suspected to condense into dust grains in the diffuse ISM, but the compounds formed from this process are unknown.
Here, we use the laboratory cross sections from \citet{Kortright:2000:PhRvB} and \citet{Lee:2009:ApJ} to model the absorption and scattering profiles of metallic Fe, and the crystalline compounds fayalite (Fe$_2$SiO$_4$), ferrous sulfate (FeSO$_4$), hematite ($\alpha$-Fe$_2$O$_3$), and lepidocrocite ($\gamma$-FeOOH), 
which have oxidation states ranging from Fe$^{0}$ to Fe$^{3+}$. 
We find that the observed Fe $L$-shell features are systematically offset in energy from the laboratory measurements. 
An examination of over two dozen published measurements of Fe $L$-shell absorption finds a $1-2$~eV scatter in energy positions of the $L$-shell features. 
Motivated by this, we fit for the best energy-scale shift simultaneously with the fine structure of the Fe $L$-shell extinction cross sections. 
Hematite and lepidocrocite provide the best fits ($\approx +1.1$~eV shift), followed by fayalite ($\approx +1.8$~eV shift). 
However, fayalite is disfavored, based on the implied abundances and knowledge of ISM silicates gained by infrared astronomical observations and meteoritic studies. We conclude that iron oxides in the Fe$^{3+}$ oxidation state are good candidates for Fe-bearing dust. To verify this, new absolute photoabsorption measurements are needed on an energy scale accurate to better than 0.2 eV.

\end{abstract}

\section{Introduction} \label{sec:intro}

Determining the composition, size distribution, and spatial distribution of iron (Fe)-bearing dust in the neutral interstellar medium (ISM) is important for understanding the history of stellar nucleosynthesis, elemental abundances in our galaxy, and stellar life cycles \citep{Dwek:1992:ARAnA, Jenkins:2009:ApJ}.  Furthermore, knowledge of the Fe-bearing dust is needed to calibrate observations of the cosmic microwave background by constraining the sources of the contaminating anomalous microwave emission attributed to dust \citep{Hoang:2016b:ApJ,Hensley:2017:ApJ836}.  Fe-bearing dust also contributes to the formation of rocky planets \citep{Ishii:2018:PNAS,Westphal:2019:ApJ}.  Ultraviolet (UV) and optical absorption studies find over $90-99\%$ of the Fe in the ISM is locked up in the solid phase, namely in interstellar dust grains \citep{Savage:1996:ARA, Jenkins:2009:ApJ}. However, those studies suffer from uncertainties in the reference abundance tables \citep{Jenkins:2009:ApJ}.
More importantly, they are unable to constrain the identity of the solid-phase Fe-bearing compounds due to the lack of dust spectral features in UV or optical \citep{Savage:1996:ARA, Jenkins:2009:ApJ, Voshchinnikov:2010:AnA}.

High resolution X-ray spectroscopy is a more powerful technique for studying Fe in the neutral ISM.  Gas-phase Fe and solid-phase Fe-bearing dust grains with radii $\lesssim 1~\mu$m are semitransparent to X-rays \citep{Wilms:2000:ApJ}.  As a consequence, high resolution X-ray spectroscopy is the sole technique that can be used to study solid phase Fe in 
the neutral diffuse ISM,
for atomic hydrogen column densities of $N({\mathrm H}) = 10^{21}-10^{23}$~cm$^{-2}$.  Using high resolution X-ray spectra of bright Galactic X-ray binaries as background light sources, Fe $L$-shell absorption and scattering features can be used to identify the abundances, ionization states, and chemical composition of the ISM.  There have been numerous past studies of Fe in the diffuse ISM \citep[e.g.,][]{Paerels:2001:ApJ, Schulz:2002:ApJ, Takei:2002:ApJ, deVries:2003:AnA, Juett:2004:ApJ, Ueda:2005:ApJ, Yao:2006:ApJ, Juett:2006:ApJ, deVries:2009:AnA, Lee:2009:ApJ, Hanke:2009:ApJ, Pinto:2010:AnA, Garcia:2011:ApJL, Costantini:2012:AnA, Pinto:2013:AnA, Gatuzz:2013:ApJ, Gatuzz:2015:ApJ, Schulz:2016:ApJ, Joachimi:2016:MNRAS, Zeegers:2017:AA, Zeegers:2019:AnA, Rogantini:2018:AA, Rogantini:2020:AnA, Psaradaki:2020:AA, Psaradaki:2023:AA, Gatuzz:2020:MNRAS, Gatuzz:2021:MNRAS}.  Many have primarily used the metallic Fe $L$-shell absorption cross section of \citet{Kortright:2000:PhRvB} for all forms of solid-phase Fe, due to a lack of data for other solid Fe-bearing compounds.  Most of these past studies have also approximated the extinction cross section with only the absorption component, leaving out the important effects of dust scattering \citep{Corrales:2016:MNRAS, Schulz:2016:ApJ, Hoffman:2016:ApJ, Zeegers:2017:AA}.  

Here, we present a new study of Fe $L$-shell extinction in the diffuse ISM.  This work improves on the current state of knowledge by taking advantage of the laboratory data of \citet{Lee:2009:ApJ} for four different crystalline Fe-bearing minerals that are candidates for forming ISM dust (fayalite [Fe$_2$SiO$_4$], ferrous sulfate [FeSO$_4$], hematite [$\alpha$-Fe$_2$O$_3$], and lepidocrocite [$\gamma$-FeOOH]). We also use the metallic Fe data of \citet{Kortright:2000:PhRvB}, as was first used for astrophysical studies by \citet{Lee:2001:ApJL}.  We model both the absorption and scattering components of the extinction cross section for the \citet{Lee:2009:ApJ} data for the first time.  We improve upon past studies by adding the experimentally benchmarked theoretical photoabsorption cross sections for Fe~{\sc i}-{\sc iv} \citep{Martins:2006:JPhB, Schippers:2017:ApJ, Beerwerth:2019:ApJ, Schippers:2021:ApJ} to our extinction model.  We also determine the gas-phase neon (Ne) and oxygen (O) abundances in neutral and ionic forms and use Ne and O as benchmarks for measuring the relative abundances and depletion of Fe in the diffuse ISM. Chemically inert Ne is not expected to be depleted in the ISM, while O appears to be depleted by only $7-20\%$ in the diffuse ISM \citep{Jenkins:2009:ApJ, Psaradaki:2020:AA, Psaradaki:2023:AA}.  

For this work, we have analyzed observations of the diffuse ISM along the lines of sight to the black hole X-ray binaries (XRBs) Cygnus~X-1 and GX~339-4.  For our analysis, we choose to focus on high signal-to-noise archival data sets from the Reflection Grating Spectrometer \cite[RGS;][]{denHerder:2001:AnA} on the \textit{X-ray Multi-Mirror Mission Newton} (\textit{XMM-Newton}).  In the energy range for Fe $L$-shell extinction, RGS data provide $10-100$ times more effective area\footnote{https://xmm-tools.cosmos.esa.int/external/xmm\_user\_support/documentation/uhb/rgsaeff.html} than that of the \textit{Chandra X-ray Observatory} gratings\footnote{https://cxc.harvard.edu/proposer/POG/pdf/MPOG.pdf}, mitigating the slightly lower RGS spectral resolution. Consequently, at these energies, some authors choose to rebin \textit{Chandra} data to achieve a significant signal-to-noise ratio \citep[e.g.,][]{Gatuzz:2015:ApJ, Westphal:2019:ApJ}, effectively reducing the spectral resolution to less than that of RGS.

This paper is organized as follows.  Section~\ref{sec:Iron} reviews the properties and current state of knowledge of Fe in the diffuse ISM.  Section~\ref{sec:ModelingFeLShell} describes the Fe $L$-shell extinction models used here and the gas- and solid-phase data incorporated into the models.  Section~\ref{sec:Observations} describes the sources, the X-ray observations, and the RGS data reduction to generate spectra.  Section~\ref{sec:SpectralAnalysis} presents the results of our spectral analysis.  Section~\ref{sec:Discussion} discusses the astrophysical implications of our findings.  A summary and conclusions are given in Section~\ref{sec:Conclusions}.

\section{Iron in the Diffuse ISM}
\label{sec:Iron}

The interstellar absorption signatures imprinted upon X-ray spectra of distant Galactic XRBs represent the total attenuation across many kpc of interstellar material, and thus samples all the known 
phases of the ISM \citep{Draine:2011:Book}: the cold neutral medium (CNM, $T \sim 100-1000$~K and $n_{\rm H} \sim 30$~cm$^{-3}$), the warm neutral medium (WNM, $T \sim 5000$~K and $n_{\rm H} \sim 0.6$~cm$^{-3}$), and the warm ionized medium (WIM, $T \sim 10^4$~K and $n_{\rm H} \sim 0.1$~cm$^{-3}$).
Here, $T$ is the gas temperature and $n_{\rm H}$ is the hydrogen nuclei number density.
In all three phases, we expect the majority of gas-phase Fe to be singly ionized (Fe~{\sc ii}). 
In the CNM, which makes up $\sim 1\%$ of the local ISM, UV and optical observations find the gas-phase ratio of Fe~{\sc ii} : Fe~{\sc i} $\sim 1.0:0.001$ \citep{Snow:2002:ApJ, Welty:2003:ApJS, Miller:2007:ApJ, Jensen:2007:ApJ, Jenkins:2009:ApJ}. This is because the ionization potential for Fe~{\sc i} at 7.8~eV is below the Lyman limit of 13.6~eV (912~\AA), above which the interstellar radiation field (ISRF) is heavily absorbed by interstellar atomic H \citep{Snow:2006:ARAA}. Thus Fe~{\sc i} is quickly ionized by the ISRF. Fe~{\sc ii}  has an ionization potential of 16.18~eV (738~\AA), above the Lyman limit, making the abundance of Fe~{\sc iii} generally negligible except near intense photoionization sources. 
The WNM has a filling factor in the Galactic disk of $\sim 40\%$ \citep{Draine:2011:Book}. It is heated, in part, by photoelectrons from dust, but the gas-phase temperature is still too low to form Fe~{\sc iii}, making Fe~{\sc ii} the dominant ion \citep{Jensen:2007:ApJ, Bryans:2009:ApJ}. 
The WIM contains $\sim 16\%$ of the baryonic mass of the Galaxy, and about half of it lies within 500~pc of the Galactic disk midplane \citep{Draine:2011:Book}. In the WIM, UV and optical observations find a gas-phase ratio of  Fe~{\sc iii} : Fe~{\sc ii} : Fe~{\sc i} ratio of $\sim 0.25 : 1.0 : 0.001$ \citep{Snow:1976:ApJ, Snow:1977:ApJ, Welty:2003:ApJS, Jenkins:2009:ApJ}. 
Thus, in these three phases of the ISM, Fe~{\sc ii} is the only gas-phase interstellar Fe species expected to produce an appreciable X-ray absorption signature. 

Studies comparing the gas-phase ISM Fe abundances to solar abundances demonstrate that the majority of Fe is depleted relative to solar and hence must have been incorporated into dust grains, which do not produce distinctive solid-Fe spectroscopic features in the UV or optical. 
The amount of Fe suspected to remain in the gas phase varies depending on the ISM temperature and location in the Galactic disk, ranging from $\sim 0.5\%$ solar abundance in cold clouds to $\sim 15-20\%$ solar abundance in the warm Milky Way halo \citep{Savage:1996:ARA, Jones:2000:JGR}. In a study of the atomic absorption from over 200 sight-lines within 3~pc of the Milky Way disk, \citet{Jenkins:2009:ApJ} finds a maximum of 10\% of interstellar Fe in the gas phase, depleting further as other elements also disappear from the gas phase. 
Thus to achieve a full inventory for the state of interstellar Fe, we must focus  predominantly on identifying the Fe-bearing compounds that make up interstellar dust, which account for the majority of the X-ray attenuation in the Fe $L$-shell photoabsorption region. 

The amount of interstellar Fe in dust grains is tied to their production by stellar nucleosynthesis and the balance between their growth and destruction rates \citep{Dwek:1998:ApJ, Nozawa:2003:ApJ, Ferrarotti:2006:AnA, Nozawa:2007:ApJ, Zhukovska:2008:AnA, Zhukovska:2018:ApJ}. Fe is predominantly produced by Type-Ia supernovae (SNe), with some contributions from core-collapse SNe and asymptotic giant branch (AGB) stars \citep{Dwek:2016:ApJ}. Type-Ia SNe have explosion energies and expansion velocities similar to core-collapse SNe, but the temperature and density of Type-Ia SNe ejecta decrease more rapidly, so there is less time available in Type-Ia SNe ejecta for the formation of molecules and grains.  Dust growth in Type-Ia SNe ejecta is also impeded by radioactive $^{56}$Ni, which produces gamma rays and fast electrons that are effective at dissociating molecular precursors of dust \citep{Nozawa:2011:ApJ}.  Furthermore, in the ejecta of Type-Ia SNe, there is no observational evidence to date for significant dust formation with a mass comparable to that observed in core-collapse SNe \citep{Gomez:2012:MNRAS}.  So it is puzzling that diffuse ISM observations of gas-phase Fe find that $90-99$\% of interstellar Fe is locked up in dust. 
This implies that much of the Fe-bearing solids in the ISM must form and grow in the diffuse ISM through a mechanism yet to be determined.

Recent studies of Fe-bearing ISM solids have made some progress in identifying their composition.  \citet{Kemper:2004:ApJ} studied $10~\mu{\rm m}$ absorption towards the Galactic Center. Their findings suggests that interstellar silicate dust consists by mass of $\sim 85$\% olivine (Mg$_{2x}$Fe$_{2(1-x)}$SiO$_4$) and $\sim 15$\% pyroxene (Mg$_{x}$Fe$_{1-x}$SiO$_3$). A broader analysis of the 10 and 20~$\mu{\rm m}$ features by \citet{Min:2007:AA} suggests equal olivine and pyroxene abundances, by mass, and that Mg-rich silicates with Mg/(Fe+Mg)$\approx 0.9$ provide a better fit to the 10 and 20~$\mu{\rm m}$ features \citep{Min:2007:AA}. This ratio aligns with the observed relative Mg and Fe abundances in GEMS\footnote{While the relative Mg and Fe abundances in GEMS appear close to that observed in the ISM, the relative O/Si and (Mg+Fe)/Si a\-bun\-dan\-ces are not \citep{Min:2007:AA}.}  \citep[Glass with Embedded Metals and Sulfides, thought to be remnants of presolar dust grains that have not been destroyed or reworked by planetary formation processes;][]{Bradley:1994:Science, Altobelli:2016:Science, Ishii:2018:PNAS}.
Given that the solar and ISM abundances of Mg and Fe seem to be approximately equal \citep{Wilms:2000:ApJ}, it is thought that the majority of interstellar Fe is not incorporated in silicates but rather in compounds such as iron oxides, ferrous sulfates, or metallic iron \citep{Henning:1995:AnAS, Jones:2013:AnA, Kohler:2014:AnA}.  

It has been hypothesized that much of the Fe in the ISM might be in the form of Fe-rich nanoparticles \citep{Hoang:2016b:ApJ, Gioannini:2017:MNRAS}.  Studies from \citet{Zhukovska:2018:ApJ} used hydrodynamical simulations for the evolution of giant molecular clouds and analytical expressions for the timescales of dust growth and destruction in order to match the depletion rates of interstellar Fe observed across different phases of the ISM.  They found that a population of $1-10$~nm sized metallic iron nanoparticles could reproduce the observed gas-phase Fe abundance profiles in the WNM, provided that 70\% of the nanoparticles were incorporated as inclusions in larger silicate dust grains; otherwise, the destruction efficiency for metallic Fe dust would have to be five times less than that for silicates. \citet{Choban:2022:MNRAS} came to similar conclusions when incorporating prescriptions for dust growth and destruction into FIRE-2 simulations\footnote{See the FIRE (Feedback in Realistic Environments) collaboration, http://fire.northwestern.edu} -- namely, they require Fe nanoparticles in order to describe the observed ISM depletion rates. Over the course of their 2~Gyr simulations, bare metallic Fe grains accounted for $\leq 10\%$ of the total ISM dust mass, while it was assumed that 70\% of the total mass of Fe nanoparticles was incorporated into silicates.  The sources of these potential Fe nanoparticles remain uncertain, as microgravity experiments of gas-phase Fe condensation in ISM-like environments show no evidence that gas-phase Fe can spontaneously condense into metallic particles \citep{Kimura:2017:SciAdv}.  Fe inclusions might instead form spontaneously through thermal annealing or in the process of silicate condensation  \citep{Davoisne:2006:AnA, Matsuno:2021:ApJ}.

Distinguishing among Fe bearing minerals in the ISM is also important for other areas of astrophysics, as Fe in dust can have effects on the physical conditions of the ISM. Whether free-flying or bound within a larger grain of different composition, metallic Fe nanoparticles can affect photoelectic heating in the ISM, can align dust grains along ambient interstellar magnetic fields, and can emit at sub-millimeter or millimeter wavelengths  \citep{Draine:2012:ApJ, Draine:2013:ApJ, Hoang:2016b:ApJ, Hensley:2017:ApJ834}.  A suspected population of free metallic Fe nanoparticles has  been invoked to explain the source of anomalous microwave emission (AME) observed at $\sim 30$~GHz \citep{Hensley:2017:ApJ834}.  However, \citet{Hensley:2017:ApJ836} also showed that Fe nanoparticles cannot fully account for the observed AME without violating other observed infrared (IR) and extinction properties of interstellar dust.  Thus, the existence of metallic Fe nanoparticles in the ISM is still a subject of debate.

Another question is the role of interstellar dust in contributing to planet formation.  Interstellar dust certainly delivers metals to protostellar nebulae, but it is yet to be determined how much of that dust survives processing in a protoplanetary disk and contributes to planetesimal formation. 
Suspected interstellar and ``presolar'' grains, including GEMS, are identified in meteorites via their isotopic anomalies relative to the solar system \citep{Zinner:1987:Nature, Nittler:2003:EPSL}. Interstellar dust is more likely to survive in the outer regions of planetary systems, leading researchers to consider some cometary dust as a potential interstellar analog \citep{Brownlee:2006:Science}. In situ collection of dust in the solar system has also yielded suspected interstellar particles based on their isotopic ratios \citep{Messenger:2000:Nature} and measured impact velocities in the collection medium \citep{Westphal:2014:Science}. 
In all of these cases, Fe appears in trace amounts. When it is observed, Fe is found to be sequestered into Fe-Ni metals and Fe-rich sulfides \citep{Hanner:2010:Springer, Mann:2010:Springer, Bradley:2010:Springer}.
Confirming whether or not these particles collected in our Solar System are truly representative of unprocessed interstellar dust requires direct measurements of solid-phase interstellar metals. High resolution absorption spectroscopy of photoabsorption structure near the photoelectric absorption edge, in the X-ray band, provides such an opportunity \citep{Lee:2005:ApJ}.

\section{Modeling Iron \textit{L}-Shell Spectroscopy in the Diffuse ISM}
\label{sec:ModelingFeLShell}

X-ray studies of Fe in the diffuse ISM focus primarily on the Fe $L$-shell features in the $0.70-0.73$~keV ($16.9-17.7$~\AA) bandpass.  Past X-ray studies of Fe in the diffuse ISM have been hindered by the lack of reliable publicly available models for Fe $L$-shell extinction by gas- and solid-phase Fe. 
We discuss this further in Sec.~\ref{subsec:Comparison}.  The field has focused on the Fe $L$ shell, in part, because the Fe $K$-shell features are more challenging to study in the ISM.  They have a much smaller absorption cross section and can only be detected for very high hydrogen column densities of $N({\rm H}) > 10^{22}$~cm$^2$ \citep{Rogantini:2018:AA}.  

In this section, we briefly review the physics of the Fe $L$-shell features. We then go on to describe the extinction models that we have updated with state-of-the-art data for gas- and solid-phase Fe and that we use for the present study.   We begin with the simpler gas-phase Fe and then describe solid-phase Fe. 

\subsection{Gas-Phase Fe}
\label{subsec:Gas-Phase}

Near neutral Fe~{\sc i}, {\sc ii}, {\sc iii}, and {\sc iv} [Fe$^{(0,1,2,3)+}$, respectively] exhibit strong resonant photoabsorption between $0.70-0.73$~keV \citep{Martins:2006:JPhB,Schippers:2017:ApJ,Beerwerth:2019:ApJ,Schippers:2021:ApJ}.  This is due to their atomic structure, with a filled $L$-shell ($2s^2 2p^6$) and a partly filled $M$-shell (a filled $3s^2 3p^6$ and partly filled $3d$ subshell).  The strongest X-ray absorption features are due to $2p \to 3d$ dipole-allowed photoexcitations.  These bound-bound transitions form $L$-shell-hole states that have insignificant fluorescence yields and predominantly autoionize \citep{Krause:1979:JPCRD,Kaastra:1993:AnASS}.  Hence, the $L$-shell photoionization and photoabsorption cross section are nearly identical for these ions.  The strongest spectral feature is due to $2p_{3/2} \to 3d$ photoabsorption ($L_3$). At slightly higher energies and less strong is the $2p_{1/2} \to 3d$ photoabsorption feature ($L_2$).  Going to higher photon energies (shorter wavelengths), the $2p \to nd$ photoexcitations to higher Rydberg levels $n$ blend together and remove any edge-like or step-like structure in the photoabsorption cross section at the direct ionization threshold, resulting in a smooth behavior above $\approx 0.73$~keV (below $\approx 17$~\AA).  

In this work, we use the experimentally benchmarked theoretical photoabsorption cross sections for Fe~{\sc i} \citep{Martins:2006:JPhB}, Fe~{\sc ii} \citep{Schippers:2017:ApJ}, Fe~{\sc iii} \citep{Schippers:2021:ApJ}, and Fe~{\sc iv} \citep{Beerwerth:2019:ApJ}. The uncertainty of the calculated resonance positions is $\sim 2-3$~eV. This is typical for calculations of transitions in the soft X-ray regime for multi-electron systems using the Hartree-Fock method and is the consequence mostly of neglecting electron correlation effects \citep{Martins:2001:JPB, Martins:2006:JPhB, Hirsch:2012:PRA}.  Smaller uncertainties can be obtained using methods that include the electron-electron interaction, such as the multiconfiguration Dirac-Fock approach, but these methods can rapidly become computationally prohibitive for complex multi-electron systems, such as the 3d metal atoms with many open shells.  For these reasons, for this work we have shifted the theoretical photoabsorption cross sections to best match the experimental data by $-2.7$, $-2.8$, $-2.5$, and $-2.2$~eV, for Fe~{\sc i}, {\sc ii}, {\sc iii}, and {\sc iv}, respectively.  The resulting energy scales are estimated to have an accuracy of $\sim 0.2$~eV, based on calibration of the energy scale of the experimental results that have been used to benchmark the theoretical calculations \cite[for details see][]{Schippers:2017:ApJ, Schippers:2021:ApJ}. These specific energy shifts were only applied to the gas-phase Fe~{\sc i}-{\sc iv} and not to any of the solid-phase Fe.  

Our modeling work here builds on the model of \citet{Gatuzz:2015:ApJ}, \ismabs, to study gas-phase X-ray extinction in the diffuse ISM.  \ismabs\ is designed for high-resolution X-ray spectroscopy and meant to be used as a local model in conjunction with the X-ray spectral fitting package {\tt XSPEC} \citep{Arnaud:1996:ASPC}.  It includes the gas-phase species H~{\sc i} [H$^{0+}$], He~{\sc i,ii} [He$^{(0,1)+}$], C~{\sc i,ii,iii} [C$^{(0,1,2)+}$], N~{\sc i,ii,iii} [N$^{(0,1,2)+}$], O~{\sc i,ii,iii} [O$^{(0,1,2)+}$], Ne~{\sc i,ii,iii} [Ne$^{(0,1,2)+}$], Mg~{\sc i,ii,iii} [Mg$^{(0,1,2)+}$], Si~{\sc i,ii,iii} [Si$^{(0,1,2)+}$], S~{\sc i,ii,iii} [S$^{(0,1,2)+}$], and Ar~{\sc i,ii,iii} [Ar$^{(0,1,2)+}$].  The only non-gas-phase species included is metallic Fe, for which \ismabs\ uses the photoabsorption data of \citet{Kortright:2000:PhRvB}. However, \citet{Gatuzz:2015:ApJ} did not include the scattering portion of the metallic Fe extinction cross section and they also blue shifted the photoabsorption cross section by 1.6 eV to best fit the astronomical observations.

For this work, we did not use the metallic Fe component of \ismabs.  Instead, we incorporated the data of \citet{Kortright:2000:PhRvB} into our model for solid-phase extinction, which includes the scattering component, as is described in more detail below.  In addition, we have modified \ismabs\ to include the energy-shifted theoretical Fe~{\sc i}-{\sc iv} gas photoabsorption data described above.

\subsection{Solid-Phase Fe}
\label{subsec:SolidPhaseFe}

When ionizing photons penetrate dust particles, the resulting absorption spectrum differs from that of the gas phase for the same element. First, the energy levels of the atom can be shifted by both the ionic bonds formed and the coordination of the atom within the solid \citep[e.g.,][]{vanAken:2002:PCM}. The shape of the solid-phase photoabsorption cross section can be further altered as the probability wave of the ejected photoelectron scatters off nearby atoms, undergoing quantum interference that leads to oscillatory variations in the absorption cross section \citep[][]{Woo:1995:ApJL, Lee:2005:ApJ}. Together, the resulting shape of the absorption cross section is broadly referred to as X-ray absorption fine structure (XAFS).\footnote{This may also be referred to as near edge X-ray absorption fine structure (NEXAFS), X-ray absorption near-edge structure (XANES), or extended X-ray absorption fine structure (EXAFS). We use XAFS as a generic catch-all term for these effects.} In analyzing high-resolution spectra obtained from Galactic X-ray binaries, XAFS provide a means to differentiate among gas and various solid phase-species of interstellar elements.

\subsubsection{Photoabsorption Data}

Fe-bearing solids exhibit $L_3$ and $L_2$ photoabsorption features that are similar to gas-phase Fe.  In metallic Fe, the Fe $L_3$ and $L_2$ photoabsorption features arise, respectively, from $2p_{3/2}$ and $2p_{1/2}$ electrons undergoing dipole-allowed transitions to unoccupied $d$-like orbitals near the Fermi level of the solid metal lattice \citep[e.g.,][]{Mott:1949:PPSA,Miedema:2013:JESRP}.  Similar transitions occur in other Fe-bearing compounds, though the energies and magnitudes of these $L_3$ and $L_2$ features can shift compared to that for metallic Fe and gas-phase near-neutral Fe \citep{Crocombette:1995:PRB, vanAken:2002:PCM, Peak:2012:EST}.  In compounds, electrons are typically removed (i.e., oxidized) from the Fe atoms to fill the outermost shell of other atoms within the solid.  For example, in fayalite (Fe$_2$SiO$_4$) and ferrous sulfate (FeSO$_4$) the Fe is ferrous (Fe$^{2+}$), while in hematite ($\alpha$-Fe$_2$O$_3$) and lepidocrocite ($\gamma$-FeOOH) it is ferric (Fe$^{3+}$).  The relevant properties of these compounds are summarized in Table~\ref{tab:compounds}, which also gives the corresponding Fe nuclei column density $N_{\rm Fe}$, determined from the stoichiometry of each compound.  The chemical shift of $\sim 1$~eV between the peaks in the photoabsorption cross section for ferrous and ferric states \citep{vanAken:2002:PCM} is readily detectable with the spectral resolution of the \textit{XMM-Newton} and \textit{Chandra} grating spectra (discuss in more detail below and in Section~\ref{sec:XrayObservations}).  

\begin{deluxetable*}{llccc}
\tablecaption{Compound properties assumed for this work. \label{tab:compounds}}
\tablewidth{0pt}
\tablehead{
\colhead{Compound} & \colhead{Formula} & \colhead{Effective} & \colhead{Density} & \colhead{$N_{\rm Fe}$} \\
& & \colhead{charge} & & \\
 & & \colhead{state} & \colhead{(g cm$^{-3}$)} & \colhead{$\left( 10^{17}~{\rm atoms} / 10^{-4}~{\rm g} \right)$}
}
\startdata
    Metallic Iron & Fe & Fe$^{0+}$  & 7.6\tablenotemark{a} & 10.78 \\ 
    Fayalite & Fe$_2$SiO$_4$ & Fe$^{2+}$ & 4.39\tablenotemark{a} & 5.91 \\ 
    Ferrous Sulfate & FeSO$_4$ & Fe$^{2+}$ & 3.65\tablenotemark{b} & 3.96 \\ %
    Hematite & $\alpha$-Fe$_2$O$_3$ & Fe$^{3+}$ & 5.3\tablenotemark{a} & 7.54 \\ 
    Lepidocrocite & $\gamma$-FeOOH & Fe$^{3+}$ & 4.0\tablenotemark{a} & 6.78 \\ 
\enddata
\tablenotetext{a}{Encyclopedia of Minerals, 2nd edition \citep{EncMinerals}}
\tablenotemark{b}{CRC Handbook of Chemistry and Physics \citep{CRCHandbook}}
\end{deluxetable*}

Spectroscopically, we require accurate photoabsorption data to distinguish the various condensed matter forms of Fe ions from one another, and from their gas-phase counterparts. Specifically, what are needed are absolute cross sections on energy scales with an uncertainty much smaller than the accuracy of the X-ray energy scale for the \textit{XMM-Newton} and \textit{Chandra} grating spectrometers. Reliable data exist for gas-phase Fe, as described above.  The situation is not the same for solid-phase Fe.  Theoretical cross-section data for solid-phase Fe $L$-shell photoabsorption are beyond current computational capabilities \citep[e.g.,][]{vanAken:2002:PCM, Miedema:2013:JESRP}.  In addition, the vast majority of laboratory measurements have reported only relative cross section scales \citep{Miedema:2013:JESRP} and are not suitable for astrophysical modeling.  

The purpose of this work is to utilize the absolute cross sections for Fe $L$-shell photoabsorption of \citet{Lee:2009:ApJ} for the crystalline minerals fayalite, ferrous sulfate, hematite, and lepidocrocite, which are predicted to be possible candidates for dust in the ISM. 
Prior to this, the only similar available cross sections were the metallic Fe results of \citet{Kortright:2000:PhRvB}. Their results have been used extensively for X-ray studies of the ISM \citep[e.g.,][]{Schulz:2002:ApJ, Juett:2006:ApJ, Hanke:2009:ApJ, Gatuzz:2015:ApJ}. For our work, we use the \citet{Kortright:2000:PhRvB} and \citet{Lee:2009:ApJ} data in our solid-phase Fe photoextinction models described below.  Absolute measurements for other Fe-bearing materials have also been reported by \citet{Westphal:2019:ApJ} and \citet{Psaradaki:2021:PhDT}. The cross-sections from the \citet{Lee:2009:ApJ} study are complementary to these data sets. \citet{Lee:2009:ApJ} include iron-oxide and Fe-S compounds not used in these aforementioned studies. In this work, we fit X-ray spectra with models that include the dust scattering component derived from the \citet{Lee:2009:ApJ} measurements  (Section~\ref{subsubsec:PhotoextinctionModel}), for the first time. 

\begin{deluxetable*}{lccccc}
\tablecaption{Experimentally measured $L_3$ peak energies in eV.\tablenotemark{a,b} \label{tab:L3}}
\tablewidth{0pt}
\tablehead{
\colhead{Reference} & \colhead{Metallic iron} & \colhead{Fayalite} & \colhead{Ferrous sulfate} & \colhead{Hematite} & \colhead{Lepidocrocite} \\
& \colhead{Fe} & \colhead{Fe$_2$SiO$_4$} & \colhead{FeSO$_4$} & \colhead{$\alpha$-Fe$_2$O$_4$} & \colhead{$\gamma$-FeOOH}
}
\startdata
\citet{Fuggle:1980:JESRP}\tablenotemark{c} & $706.7\pm1.0$ & & & & \\
\citet{Leapman:1982:PRB}        & 707.3 & & & & \\ 
\citet{Fink:1985:PRB}\tablenotemark{c}     & 706.6 & & & & \\
\citet{Krishnan:1990:UM}        & & 710.5 & & & \\ 
\citet{Krivanek:1990:UM}        & & & & 708.4 & \\ 
\citet{Kuiper:1993:PRL}         & & & & 710.0 & \\ 
\citet{Garvie:1994:AM}\tablenotemark{c}   & & & & $709.5\pm0.2$ & \\ 
\citet{Chen:1995:PRL} & 707.7 & & & & \\ 
\citet{Crocombette:1995:PRB}    & & & & 702.8 & \\ 
\citet{Schedel-Niedrig:1995:PRB}\tablenotemark{c} & & & & $711.3\pm0.5$ & \\ 
\citet{Garvie:1998:Nat}         & 707.8 & 707.7 & & & \\ 
\citet{Kurata:1998:JSEM}        & & & & 709.2 & \\ 
\citet{Berling:1999:JMMM}       & 707.5 & & & & \\ 
\citet{Gloter:2000:PRB}         & & & & 711.2 & \\ 
\citet{Kortright:2000:PhRvB}\tablenotemark{d} & 706.9 & & & & \\
\citet{Regan:2001:PRB}          & & & & 709.3 & \\ 
\citet{Mitterbauer:2003:UM}     & & & & 708.9 & \\ 
\citet{Todd:2003:GCA}\tablenotemark{c} & & & & 710.2 & \\ 
\citet{Calvert:2005:JESRP}\tablenotemark{c} & & & & 709.5 & \\ 
\citet{deGroot:2005:JPCB}       & & 707.9 & & 709.5 & \\ 
\citet{Vayssieres:2005:AdvMat}  & & & & 709.2 & \\ 
\citet{Wang:2006:UM}            & 709.0 & & & & \\ 
\citet{Braicovich:2007:PRB}     & 708.4 & & & & \\ 
\citet{Kang:2008:PRB}           & 707.3 & & & & \\ 
\citet{Chen:2009:PRB}           & & & & 709.7 & \\ 
\citet{Feldhoff:2009:JSSC}\tablenotemark{c} & 708.1 & & & 710.8 & \\ 
\citet{Lee:2009:ApJ}\tablenotemark{d} & & 707.3 & 706.4 & 708.0 & $708.3$ \\ 
\citet{Galakhov:2010:JPCC}      & 708.3 & & & & \\ 
\citet{Takagi:2010:PRB}         & 706.6 & & & & \\ 
\citet{Titov:2010:CPL}\tablenotemark{c} & & & & 709.5 & \\ 
\citet{Chen:2011:PCCP}          & 706.8 & & & & \\ 
\citet{Braun:2012:CPC}          & & & & 707.1 & \\ 
\citet{Hirsch:2012:PRB}         & 708.0 & & & \\ 
\citet{Peak:2012:EST}\tablenotemark{c} & & & & 709.5& \\ 
\\
Average measurement                  & $707.5\pm0.7$ & $708.4\pm1.2$ & & $709.1\pm1.8$ & \\
\enddata
\tablenotetext{a}{Primarily drawn from the review of \citet{Miedema:2013:JESRP} and the associated website https://anorg.chem.uu.nl/xaseels/.  We are unaware of any review covering from 2013 to the present.}
\tablenotetext{b}{Unless otherwise noted, read off graphs with an estimated reading uncertainty of approximately $\pm 0.3$~eV.}
\tablenotetext{c}{Explicitly given in the reference and, if available, the uncertainty.}
\tablenotetext{d}{Private communication.}
\end{deluxetable*}

Surveying the published relative and absolute Fe $L$ photoabsorption data \citep[most recently reviewed in][]{Miedema:2013:JESRP}, it is clear that one of the biggest uncertainties is the accuracy of the energy scales of the data.  This can be readily seen by the comparison in Table~\ref{tab:L3} for the experimentally measured peak energies of the $L_3$ feature for metallic iron, fayalite\footnote{The fayalite $L_3$ feature reported by \citet{Krishnan:1990:UM}, \citet{Garvie:1998:Nat}, and \citet{deGroot:2005:JPCB} is single peaked.  However, \citet{Lee:2009:ApJ} report an $L_3$ feature that consists of two peaks of nearly the same amplitude at 706.6 and 708.3~eV.  This may be due to the higher energy resolution of their measurements.  Here, we use the average energy of these two peaks.}, and hematite.  The published values range, respectively, from $706.7-709.0$, $707.5-710.5$, and $702.8-711.3$~eV.  These ranges are much larger than the accuracy of the wavelength scales for RGS\footnote{See Table~8 of https://xmm-tools.cosmos.esa.int/external/xmm\_user\_support/documentation/uhb/XMM\_UHB.pdf} and for the \textit{Chandra} gratings\footnote{See https://cxc.harvard.edu/proposer/POG/html/chap8.html} of $\pm 5$ and $\pm 11$~m\AA, respectively. In the Fe $L$ bandpass, these correspond to energy scale uncertainties of approximately $\pm 0.2$ and $\pm 0.4$~eV, respectively.   For lepidocrocite and ferrous sulfate, we are unaware of any measurements other than those of \citet{Lee:2009:ApJ} with which to compare. In order to account for the uncertainties in the energy scale of the laboratory data, we found that we needed to introduced an additive shift to the energy scale for our photoextinction modes to best match to the observational data, as is described in Section~\ref{subsec:SolidFeFit}.
The average and standard deviation of the measured energy position of the $L_3$ peak is reported in Table~\ref{tab:L3} merely to convey the high variance found in the literature.

\subsubsection{Photoextinction Model}
\label{subsubsec:PhotoextinctionModel}

Computing the absorption and scattering properties of a material requires knowledge of the real and imaginary parts of the index of refraction. The absorption cross sections derived via laboratory work provide the imaginary part of the optical constants; the real part can be derived via the Kramers-Kronig relations \citep{Kronig:1926:JOSA, Lucarini:2005:book}. 
We use the published photoabsorption data of \citet{Kortright:2000:PhRvB} for metallic Fe and of \citet{Lee:2009:ApJ} for crystalline fayalite, ferrous sulfate, hematite and lepidocrocite.
For each compound, we used the {\tt kkcalc} software \citep{Watts:2014:OpticsExpress} to apply the Kramers-Kroenig transformation on the Fe $L$-shell absorption data.  Table~\ref{tab:compounds} summarizes the chemical formula (i.e., stoichiometry) and mineral densities \citep{EncMinerals,CRCHandbook} that are supplied as inputs to {\tt kkcalc}, which uses that information to model photoionization cross sections of the remaining elements in the compound \citep[following][]{Henke:1993}, and to
extrapolate the absorption curve up to 500~keV \citep[following][]{Biggs:1988}. As a result, our model optical constants span the entire wavelength range covered by the {\it XMM-Newton} RGS observations and includes a rudimentary model for photoabsorption from other elements in each compound (e.g., O and Si).

With the optical constants and the Mie scattering algorithm of \citet{BHMie}, we compute the absorption and scattering cross sections for spherical grains.  Given the sub-arcminute imaging resolution of {\it XMM-Newton}, as well as the energy range and grain sizes considered, the amount of scattered light captured by the spectral extraction aperture is negligible \citep{Corrales:2016:MNRAS}. The final dust extinction cross section was computed by summing the absorption and scattering cross sections of each grain and then integrating over the grain size distribution.
We follow the ISM dust size properties derived by  \citet[][hereafter MRN]{Mathis:1977:ApJ}.

To apply the cross sections in a standard X-ray spectral fitting framework, we build on the high resolution ISM photoextinction model, {\tt ISMdust} \citep{Corrales:2016:MNRAS}, a local model designed for use in {\tt XSPEC} \citep{Arnaud:1996:ASPC}. The default version of {\tt ISMdust} uses generic extinction cross sections for graphitic and silicate dust grains, derived from the optical constants of \citet{Draine:2003:ApJ}, where Fe photoabsorption is provided solely by the silicate component. 
We generated a custom version of {\tt ISMdust}, dubbed {\tt SolidFeL}, to investigate solid Fe $L$-shell extinction from an MRN distribution of spherical dust particles.
The transmitted flux $F$ for each model is given by
\begin{equation}
    \frac{F}{F_a} = \exp\left(-\Sigma_{i} \kappa_{i} M_{{\rm d},i} \right),    
\end{equation}
where $F_{a}$ represents the flux of the gas-phase absorbed spectrum and $\kappa_{i}$ is the extinction opacity (in g$^{-1}$~cm$^{2}$) for each compound $i$ integrated over the dust grain-size distribution.  The free parameter in each model is $M_{{\rm d},i}$, which is the dust mass column (in g~cm$^{-2}$) of compound $i$. 
The {\tt SolidFeL} extinction opacities are computed following the MRN power-law slope of $-3.5$ with dust particle radii from 5~nm to $0.3~\mu {\rm m}$.

\section{Observations}
\label{sec:Observations}

\subsection{Cygnus X-1 and GX 339-4 }
\label{sec:Sources}

The Galactic
black hole X-ray binaries Cygnus X-1 and GX~339-4 make excellent background sources by which to study the composition of the line-of-sight ISM via X-ray absorption \citep[e.g.,][]{Schulz:2002:ApJ, Juett:2006:ApJ, Hanke:2009:ApJ, Lee:2009:ApJ, Gatuzz:2016:AnA, Westphal:2019:ApJ}. Their ISM column densities lie in an ideal range, large enough to produce deep Fe $L$-shell extinction features without completely extinguishing the X-ray count rate in the corresponding energy range. The broad-band X-ray spectrum of the two sources are generally described by a two-component model with a variable intensity \citep[e.g.,][]{Tanaka:1995:INBOOK,Schulz:2002:ApJ} consisting of a power-law component and a multicolor disk component. As the two sight lines are nearly orthogonal to each other, they sample the ISM in unique directions around the Galactic plane. 

\cygnusx1\ ($l_{\rm II} = 71.33^{\circ}$, $b_{\rm II} = 3.07^{\circ}$)\footnote{Galactic longitude and latitude (B1950)} was discovered at the dawn of X-ray astronomy \citep[][and reference therein]{Giacconi:1967:ApJ} and subsequently found to consist of an $\approx 21~M_\odot$ black hole and an $\approx 41~M_\odot$ O9.7~Iab  stellar companion \citep{Sota:2011:ApJS, Miller-Jones:2021:Sci} at a distance of 
$\sim 2.2$~kpc 
\citep{Ziolkowski:2005:MNRAS, Gaia:2020:Cat}. It has a binary orbit of $\approx 5.6$~days \citep{Brocksopp:1999:AnA},  an inclination of $i \approx 27^\circ$ \citep{Miller-Jones:2021:Sci}, and a radial velocity of $-2.7 \pm 3.5$~km~s$^{-1}$ \citep{Gontcharov:2006:AstL}.  
It is characterized by two distinct spectral states with a power-law photon index $\mathit{\Gamma}_{\rm hard} \sim 1.7$ and $\mathit{\Gamma}_{\rm soft} \sim 2.5$, respectively. The disk component is most manifest in the hard state and difficult to detect otherwise. This source is highly variable and displays intervals of flaring activity.

GX~339-4 ($l_{\rm II} = 339.94^{\circ}$, $b_{\rm II} = -4.33^{\circ}$) is a black hole XRB discovered in outbursts by \citet{Markert:1973:ApJ}, a state found to repeat $1-3$ times per year. Although GX~339-4 is well-studied, its physical parameters remain somewhat uncertain due to the weakness of its companion star \citep{Cowley:2002:AJ, Hynes:2004:ApJ, Zdziarski:2004:MNRAS,  Parker:2016:ApJ, Heida:2017:ApJ, Sreehari:2019:AdSpR, Zdziarski:2019:MNRAS}. The most recent near-IR study by \citet{Heida:2017:ApJ} suggest a K-type donor star and confirms a reported binary orbit of 1.76~days \citep{Hynes:2003:ApJ, Levine:2006:ATel}. \citet{Heida:2017:ApJ} estimate an inclination between $37^{\circ} - 78^{\circ}$ and radial velocity of $64 \pm 8$~ km~s$^{-1}$. They derived a binary mass function $f(M) = 1.91 \pm 0.08\ M_\odot$, a factor $\sim 3$ lower than the earlier work, and predict a black hole mass in the range of $2.3 - 9.5\ M_\odot$.  The latest distance estimate to GX~339-4 is 5~kpc \citep[Gaia DR3,][]{Gaia:2020:Cat}

\subsection{XMM-Newton RGS Data} 
\label{sec:XrayObservations}

A wealth of archival {\it XMM-Newton} observations are available for both binaries of interest. \cygnusx1\ was observed 13 times between $2004 - 2016$, using the RGS as the prime instrument, and acquired in various flux states, from which we select a subset based on its binary orbital phase, as described below. 
GX 339-4 was observed 21 times in its various flux states between $2001-2015$. We selected observation from the 2015 transition back to the hard, quiescent flux state \citep{Petrucci:2017InProceedings}.
Logs of these observations are presented in Tables~\ref{tab:cygobslog} and \ref{tab:gxobslog}, respectively.

The flux from both sources is too bright for the nominal image spectrometry with \xmm, and they are best studied using the high spectral resolution RGS \citep{denHerder:2001:AnA}. RGS consists of two detectors, RGS1 and RGS2, comprising a linear array of nine charge coupled devices (CCDs) operated in single-photon-counting and frame-transfer mode. Each detector sits at the focal plane behind an X-ray mirror / reflection grating assembly pair that disperse about half of the incident light onto the RGS detector plane. For a given X-ray source, the dispersed spectrum appears as a set of arcs, one for each spectral order. For this work we restrict our analysis to the summed first-order spectrum to maximize the signal-to-noise. RGS achieves a resolving power of $R = E/ \Delta E = 150 - 800$, limited by the $\approx$ $6^{\prime\prime}$~full width at half maximum (FWHM) spatial resolution of the telescope mirrors. At the Fe $L$-shell features of interest in this study, RGS has a combined a first-order effective area of $A_{\mathrm{eff}}\sim 110$~cm$^2$ and energy resolution of $\Delta E \approx 2.5$~eV. The RGS energy-scale calibration accuracy in the vicinity of the Fe $L$-shell features is $\pm 0.2$~eV, less than the chemical shift between the ferrous and ferric Fe $L$-shell features considered in this work. 

\begin{deluxetable*}{ccccc}
\tablecaption{\xmm\ RGS Observations of \cygnusx1\ Used in this Work \label{tab:cygobslog}}
\tablewidth{0pt}
\tablehead{
\colhead{ObsID}  & \colhead{Start Date} & \colhead{Exposure}  & \colhead{Mid MJD} & \colhead{Mid Phase} \\
\colhead{}       & \colhead{}      & \colhead{(ks)}      & \colhead{(TBD)}   & \colhead{}
}
\startdata
0745250501 & 2016 May 29 & 	144.8 & 57538.74449  & 0.324 \\
0745250601 & 2016 May 31 & 	145.4 & 57540.76372  & 0.684 \\
\enddata
\tablecomments{
Only the portion of the two observations that fell within the quadrant outside of conjunction with the O-star were considered in this work, with a total filtered exposure time of 195~ks. Here, ObsID is Observation Identification number, UTC is Coordinated Universal Time, MJC is Modified Julian Date, and TBD is Barycentric Dynamical Time.}
\end{deluxetable*}

\begin{deluxetable*}{ccc}
\tablecaption{\xmm\ RGS Observations of \gx339\ Used in this Work \label{tab:gxobslog}}
\tablewidth{0pt}
\tablehead{
\colhead{ObsID}  & \colhead{Start Date} & \colhead{Exposure}   \\
\colhead{}       & \colhead{}      & \colhead{(ks)} }
\startdata
0760646201 & 2015 Aug 28  & 18.8 \\
0760646301 & 2015 Sep 02  & 18.0  \\
0760646401 & 2015 Sep 07  & 22.5  \\
0760646501 & 2015 Sep 12  & 20.9  \\
0760646601 & 2015 Sep 17  & 48.7  \\
0760646701 & 2015 Sep 30  & 48.0  \\
\enddata
\end{deluxetable*}

\begin{figure}
\plotone{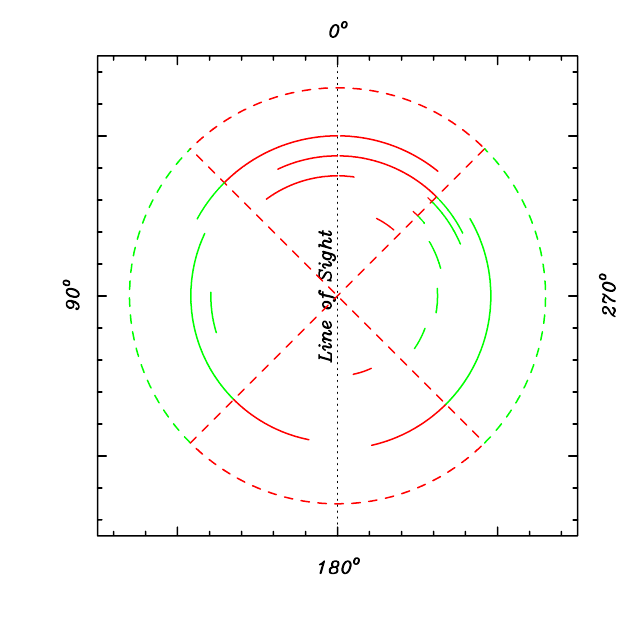}
\caption{Graphical representation of the binary phase of the \cygnusx1\ black hole during the \xmm\ RGS observations relative to the inferior conjunction of the O-star, as determined from the radial-velocity-derived ephemeris of \citet[][$P = 5.599829$~days at phase-zero epoch 41874.207 MJD]{Brocksopp:1999:AnA}. Each observation in Table~\ref{tab:cygobslog} is drawn as a solid arc in phase that covers its exposure, ordered chronological with the innermost arc for the earliest observation.  In order to avoid possible phase-dependent interaction of the X-rays from the black hole with the wind from the companion star, we only consider exposure intervals in the green quadrants in our analysis. 
\label{fig:phase}
}
\end{figure}

All observations were reprocessed using the latest calibrations and  reduced and analysed using the Science Analysis Software (SAS) version {\tt xmmsas\_20190531\_1155-18.0.0}. Nominal RGS1  and  RGS2  source and background first-order spectral files and their response matrices were generated using the SAS {\tt rgsproc} task.  This task calls {\tt rgsspectrum} to  extracted source photons falling on a rectangle aperture whose  width covers 95\% of the point spread function (PSF). The background is extracted from the adjacent regions well outside the target PSF. There are no known contaminating sources in the extracted field. Intervals of high background were automatically identified and filtered out of the event files using the default threshold applied to the background light curves. The resulting spectra from the two detectors
were combined using the SAS task {\tt rgscombine} after verifying that spectral fits to the individual components were in agreement within the measurement uncertainties. Not all RGS CCDs were active during all observations, but by merging the RGS1 and RGS2 data,  
a continuous dispersed spectrum was obtained over the energy range of interest for the Fe $L$-shell absorption study.

These data were acquired in {\tt HighEventRateWithSES} mode (i.e., SAS “Spectroscopy HER+SES" mode), which is appropriate for bright sources. Nevertheless, some of the brighter Cygnus X-1 observations suffered from pile-up, when more than one photon fell in a pixel during a readout cycle, 
distorting the spectrum.  These observations were identified and rejected using the fractional difference in flux between grating orders \citep[e.g.,][]{Costantini:2012:AnA}.

Furthermore, the intrinsic spectrum of  \cygnusx1\ may be modified due to periodic interactions of the X-ray emisison from the compact object with the wind from the binary companion. To minimize these possible spectral distortions, we restricted our analysis to data accumulated during binary phase intervals well away from the inferior or superior conjunction of the O-star, following the approach of \citet{Westphal:2019:ApJ}, using the  
the radial-velocity-derived ephemeris of \citet{Brocksopp:1999:AnA}.
This is illustrated in Figure~\ref{fig:phase}, where we highlight in green suitable portion of each observation.  
We truncated and combined the two long observations that fell mostly within the quadrants of the orbital phase perpendicular to conjunction (given in Table~\ref{tab:cygobslog}). These observations were close in time and have similar fluxes and spectra. In contrast, the orbital ephemeris for GX~339-4 is not yet well established and we used the observations selected for consistent spectral state.

After reprocessing, reducing, and filtering the data sets, eliminating piled-up observations, and applying orbital phase and/or epoch selection, a single merged source spectrum for each source was generated, along with each associated background spectrum and response matrix. 
The final Cygnus X-1 spectrum yielded a filtered exposure time of 195~ks, with a net count rate of 11.3~cps in the full RGS energy band,  
and similarly, 444~ks and 3.11~cps for GX~339-4. 
We fit the unbinned spectra using the {\tt XSPEC} spectral fitting program \citep[version 12.10.1f,][]{Arnaud:1996:ASPC}. All spectral channels are well populated around the Fe features of interest; however, for \gx339, below $0.5$~keV, the counts in some channels are $<10$. Hence, for consistency, we used C-Statistic, $C_{\rm stat}$, tests \citep{Cash:1979:ApJ} for all fits.

\section{Spectral Analysis and Results}
\label{sec:SpectralAnalysis}

\subsection{Baseline Continuum Model}

Our aim is to identify possible contributions to the ISM by the Fe-bearing compounds codified in the {\tt SolidFeL} model.
Our method is to establish the baseline continuum model for the observed spectra of \cygnusx1\ and \gx339, outlined below, and then evaluate the likely contribution to the ISM of trial Fe compounds for the fixed continuum.

The spectral continua of \cygnusx1\ and \gx339\ over a nominal $0.45-1.80$~keV RGS bandpass are generally characterized by a simple non-thermal power-law model. The only evidence for X-ray reflection is from two broad emission features that are  modeled here by two Gaussians, one centered at an energy slightly below the Fe $L$-shell photoabsorption edge, and the other at an energy just below the Ne $K$-shell edge.  Similar features have been found in spectra of other XRBs \citep{Madej:2010:MNRAS, Madej:2014:MNRAS, Psaradaki:2020:AA}.  

To account for ISM absorption in the spectra of the two sources, we used a combination of the \ismabs\ and {\tt ISMdust} models. 
With \ismabs, we accounted for contributions from C~{\sc ii}, N~{\sc i}, O~{\sc i, ii, iii}, Ne~{\sc i}, and Mg~{\sc ii}.  In the diffuse ISM, these are the dominant gas-phase species for their elements, due to the combination of their ionization potentials and the shape of the ISRF \citep{Snow:2006:ARAA}. The absorption edges of C~{\sc ii} and N~{\sc i} fall below our fitted energy range; however, they provide an important contribution to the continuum shape, and their column densities are estimated as a constant fraction of
hydrogen column density $N({\rm H})$,
using the abundances and depletion factors of \citet{Wilms:2000:ApJ}.  \citet{Gatuzz:2016:AnA}, in their analysis of both sources, found absorption by O~{\sc i} and Ne~{\sc i} and, at a much reduced level, O~{\sc ii} and O~{\sc iii} as well as Ne~{\sc ii} and Ne~{\sc iii}. In our spectra, we clearly identify O~{\sc ii} and O~{\sc iii} but neither Ne~{\sc ii} nor Ne~{\sc iii} is  evident. In this baseline continuum model, we did not include absorption by gas-phase Fe, as it has been inferred to be heavily depleted in the diffuse ISM.  We used {\tt ISMdust} to account for absorption by graphite and silicate dust grains. Following the findings of \citet{Corrales:2016:MNRAS}, we set the dust-to-hydrogen mass ratio to 1\% and the silicate-to-graphite grain-mass ratio to 60/40.

\begin{figure}[t]
{
\hfill
\hfill
\centerline{\includegraphics[width=0.5\textwidth,clip=true]{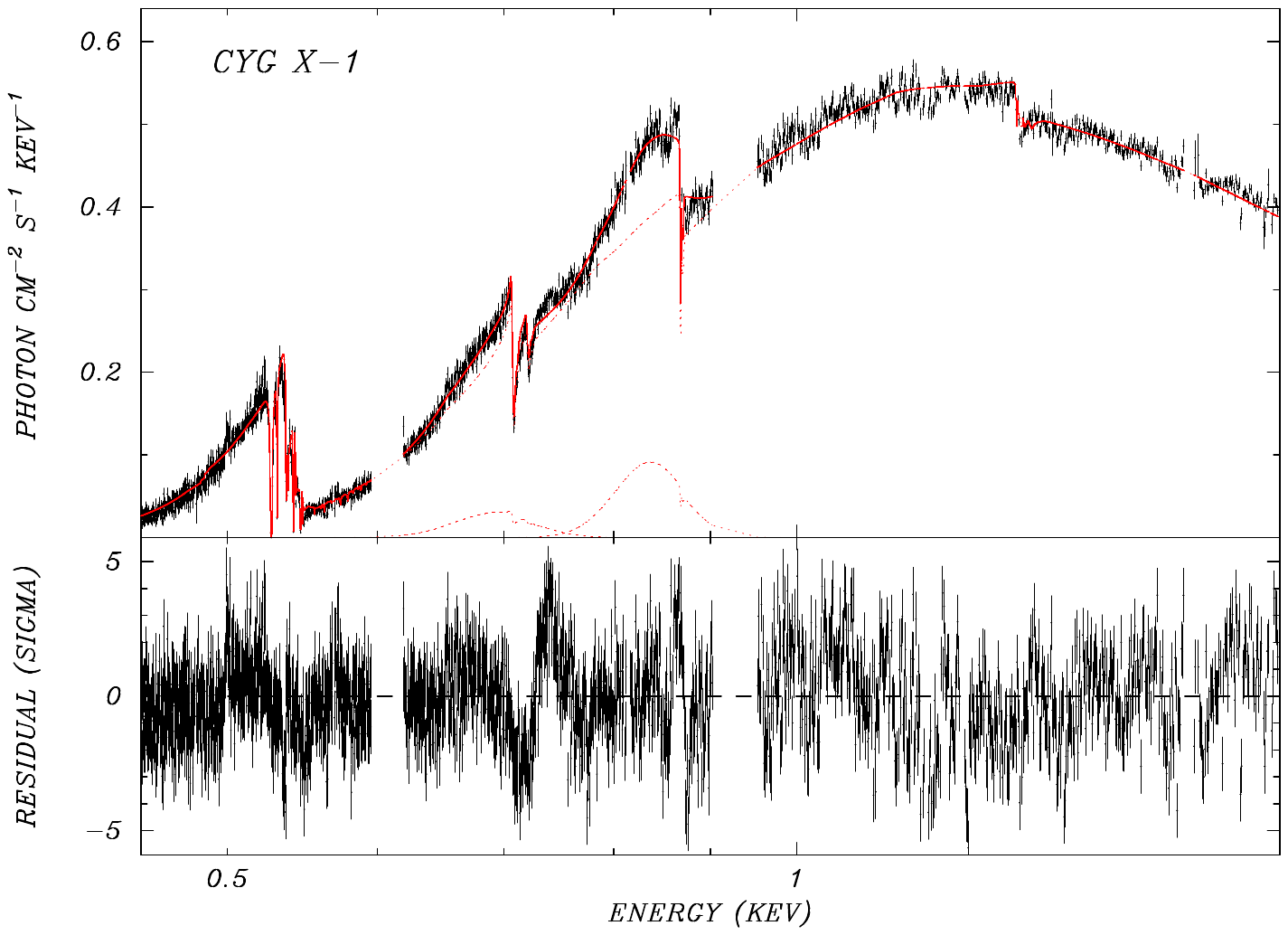}
\hfill
\includegraphics[width=0.5\textwidth, clip=true]{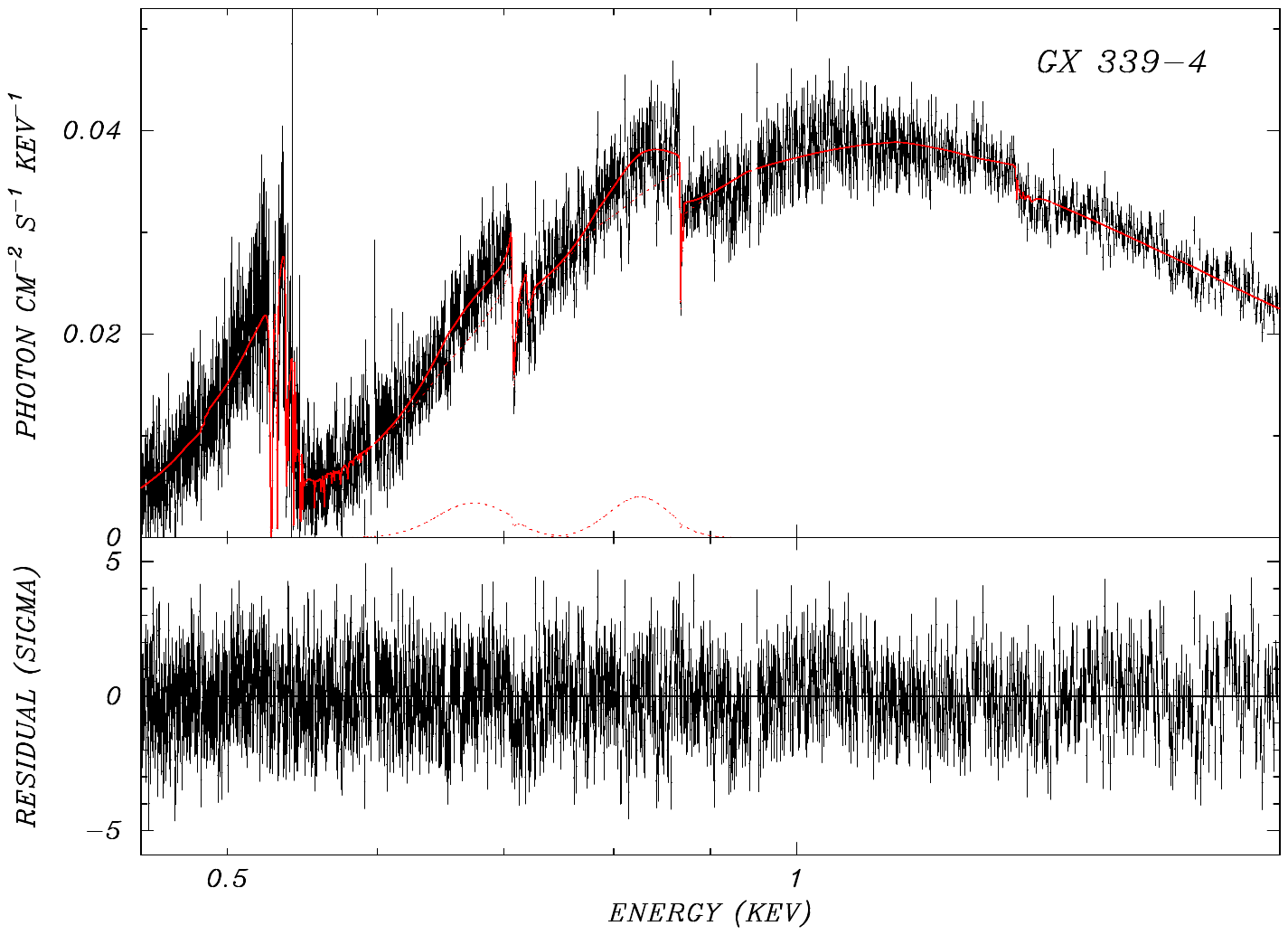}
\hfill
}
\vspace{-30pt}
\caption{\label{fig:fullCygX1}
The RGS spectra (crosses) for Cygnus~X-1 (left) and GX~339-4 (right) using the parameters given in Table~\ref{tab:specresults}. The vertical error bars show the $1\sigma$ statistical uncertainty and the horizontal the energy bin widths.  In each upper panel, the best-fit continuum model is shown by the solid lines, while the dashed lines indicate the model components. The lower panels show the residuals from the fitted model in units of $\sigma$. The gap in the Cygnus~X-1 spectrum is due to the loss of corresponding RGS CCD detectors covering those dispersed wavelengths.}
}
\end{figure}

The result of fitting the baseline continuum model (power-law, Gaussian emission features, \ismabs, and {\tt ISMdust}) is shown in Figure~\ref{fig:fullCygX1} for the two sources. Their spectral parameters are presented in Table~\ref{tab:specresults}. Both sources are in their soft state and the only spectral features from their expected surrounding warm disks are the inferred X-ray reflection features.  Previous studies have found weak spectral features from from O~{\sc vii} or {\sc viii} that is attributed to hot gas either in the ISM or local to the source.  For example, for GX 339-4, \citet{Pinto:2013:AnA} and \citet{Gatuzz:2018:MNRAS} detect hot gas, but the $N_{\rm H}$ of the hot gas is $\lesssim 2\%$ of the cold gas.  Here, we found no statistically significant spectral features for either source, although this could be due to orbital phase differences in the data used for the present analysis.  The ISM column densities and underlying continuum models are consistent with previous work \citep[e.g.,][]{Schulz:2002:ApJ, Juett:2006:ApJ, Hanke:2009:ApJ, Lee:2009:ApJ, Gatuzz:2016:AnA, Westphal:2019:ApJ}. The \texttt{ISMdust} model describes dust abundance using a mass column, $M_{\rm d}$  in units of $10^{-4}$~g~cm$^{-2}$, from which number densities are derived following an MRN power-law distribution of dust grains from 0.005 to 0.3~$\mu{\rm m}$. The silicate and graphite dust mass columns are denoted $M_{\rm sil}$ and $M_{\rm gra}$, respectively, in Table~\ref{tab:specresults}.

\begin{deluxetable*}{lccccc}[ttt]
\tablewidth{0pt}
\tablecaption{Best-fit Spectral Results for the Baseline Continuum Model\tablenotemark{a}}
\tablehead{
\colhead{} & \colhead{} & \colhead{} & \colhead{Cygnus~X-1 } & \colhead{GX~339-4}  & \colhead{} \\
\colhead{Model} & \colhead{Parameter} & \colhead{Units} & \colhead{Value} & \colhead{Value} & \colhead{Notes}
}
\startdata
Coordinates   &  ($l_{\rm II}$, $b_{\rm II}$)     &  degree &    (71.334, 3.066)     &   (338.939, -4.326)    & 	Galactic (B1950) \\
\tableline 
\ismabs\  &  \nh\              &  $10^{22}$~cm$^{-2}$ &  $0.567\pm0.004$ &  $0.49\pm0.01$       &    		 \\
          &  $N$(C~{\sc ii})   &  $10^{16}$~cm$^{-2}$ &  $68.04\pm0.04$ &  $59.0\pm0.1$        &
          $N$(C~{\sc ii}) $ = 1.2\times 10^{-4}$ \nh\ \\
          &  $N$(N~{\sc i})    &  $10^{16}$~cm$^{-2}$ &  $43.09\pm0.03$ &  $37.4\pm0.8$        & $N$(N~{\sc i})  $ = 7.6\times10^{-5}$ \nh\  \\
          &  $N$(O~{\sc i})    &  $10^{16}$~cm$^{-2}$ &  $310\pm5$    &  $273\pm19$          &    		 \\
          &  $N$(O~{\sc ii})   &  $10^{16}$~cm$^{-2}$ &  $6.4\pm2.8$    &  $18_{-11}^{+17}$    &    		 \\
          &  $N$(O~{\sc iii})  &  $10^{16}$~cm$^{-2}$ &  $5\pm2$   &  $6_{-4}^{+9}$ &    		 \\
          &  $N$(Ne~{\sc i})   &  $10^{16}$~cm$^{-2}$ &  $45\pm3$   &  $39\pm7$          &    		 \\
          &  $N$(Mg~{\sc ii})  &  $10^{16}$~cm$^{-2}$ &  $15.6\pm2.7$   &  $14\pm8$            & 		 \\
\tableline
Power-law &  $\mathit{\Gamma}$ &  unity              &  $2.36\pm0.03$                 &  $2.63\pm0.07$              &    		 \\
          &  $N_{\Gamma}$              &  arbitrary          &  $2.36\pm0.04$                 &  $0.152\pm0.006$            &   		 \\
\tableline
Gaussian  &  $E_1$             &  keV                &  $0.830\pm0.002$               &  $0.821\pm0.007$            &   	 \\ 
          &  $N_{E_1}$         &  arbitrary          &  $(7.8\pm0.5)\times10^{-2}$    &  $(2.5\pm0.7)\times10^{-3}$   & 		 \\
          &  $E_2$             &  keV                &  $0.679\pm0.003$               &  $0.664\pm0.008$            &   	 \\ 
          &  $N_{E_2}$        &  arbitrary          &  $(7.4\pm1.1)\times10^{-2}$    &  $(5.8\pm3.2)\times10^{-3}$  &              \\ 
          &  $\sigma_{1,2}$    &  keV                &  $(3.6\pm0.5)\times10^{-2}$   &  $(3.2\pm0.7)\times10^{-2}$ & $\sigma_{1,2}$ are linked	\\ 
\tableline
{\tt ISMdust}&  $M_{\rm sil}$           & $10^{-4}$~g~cm$^{-2}$ &  $0.569\pm0.004$             &  $0.494\pm0.010$            & $= 1.00357$ \nh\tablenotemark{b}  \\
             &  $M_{\rm gra}$           & $10^{-4}$~g~cm$^{-2}$ &  $0.379\pm0.003$             &  $0.329\pm0.007$           & $= 0.669$ \nh\tablenotemark{b} \\
\tableline
$C_{\rm stat}$(DoF) &          &  &   4295.58 (1814)    &  2580.84(2013)&    		 \\
\enddata
\label{tab:specresults}
\tablenotetext{a}{{\tt XSPEC} model: {\ismabs\ * {\tt ISMdust * (Pow+Gau+Gau)}}. The {\tt ISMabs} parameter $N(X)$ is the column density for species $X$.  The power-law model for the flux is of the form $F(E) = N \exp(-\mathit{\Gamma})$.  The Gaussian is of the form $F(E) = N_{E_x} \exp[- (E-E_x)^2 / (2 \sigma^2)]$. See text for details.}
\tablenotetext{b}{{\tt ISMdust} parameters $M_{\rm sil}$  and $M_{\rm gra}$  are normalized in units of $10^{-4}$~g~cm$^{-2}$ equivalent to \nh\ in units of $10^{22}$~cm$^{-2}$.}
\tablecomments{Continuum spectral fits are for the $0.45-1.8$ keV range.
Quoted uncertainties correspond to a 90\% confidence level. for two interesting parameters.
The degrees of freedom (DoF) for the fit statistic is given in parentheses.}
\end{deluxetable*}

\subsection{Fitting Solid-Phase Fe Photoabsorption}
\label{subsec:SolidFeFit}

\begin{figure}[t]
{
\hfill
\centerline{\includegraphics[width=0.5\textwidth, clip=true]{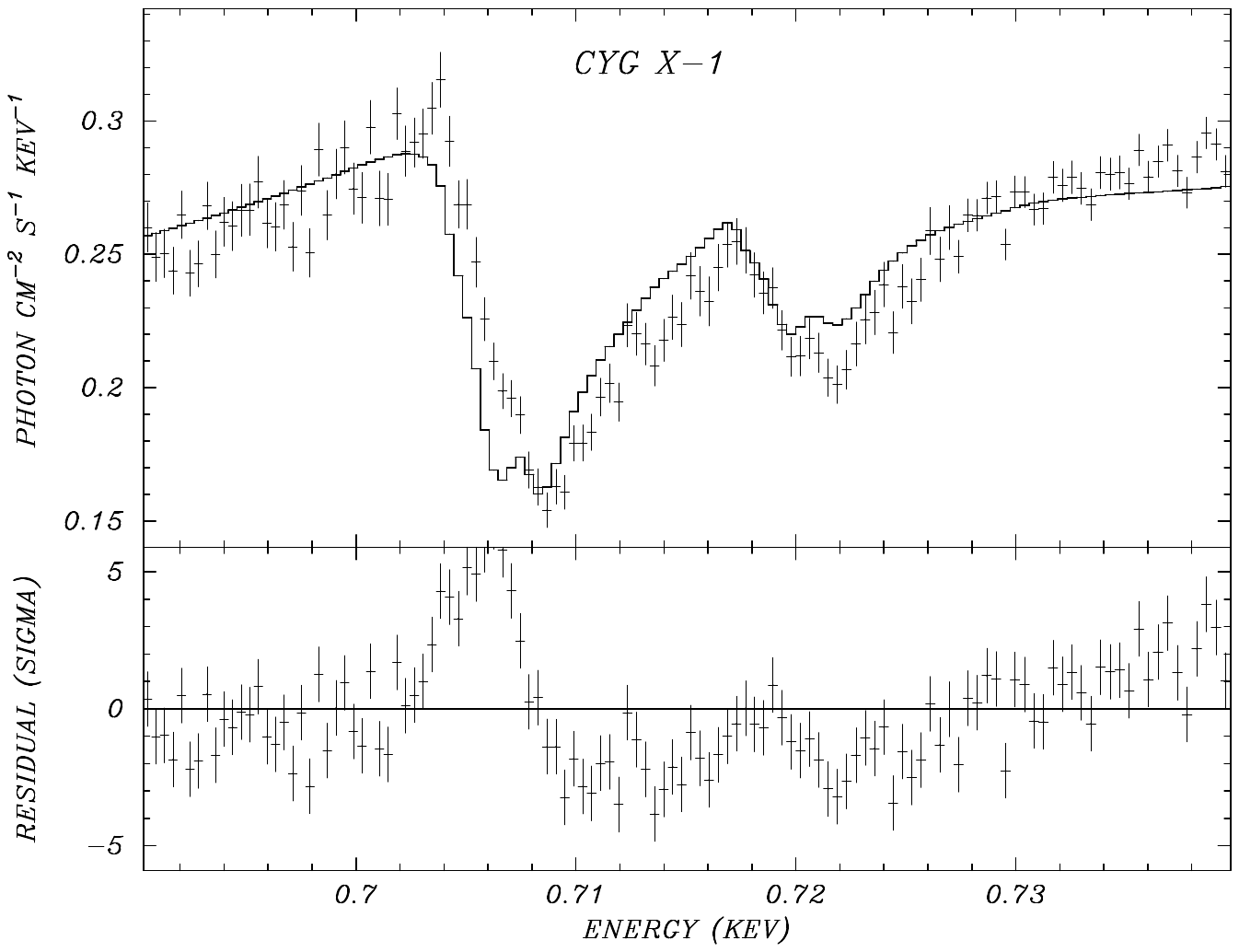}
\hfill
\includegraphics[width=0.5\textwidth, clip=true]{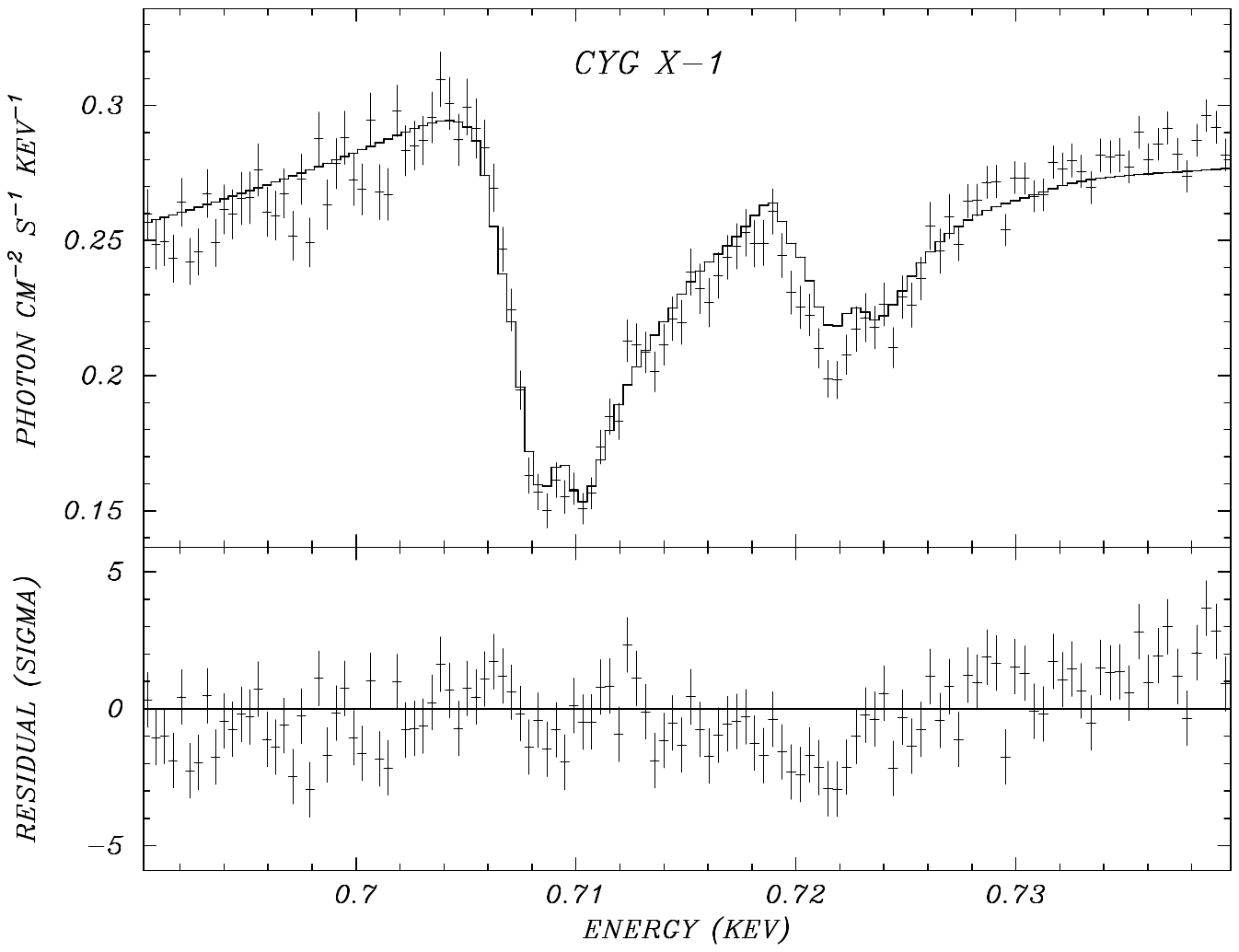}
\hfill
}}
\vspace{-30pt}
\caption{\label{fig:fayalite}
RGS spectra of the Fe feature in \cygnusx1\ fitted in the $0.69-0.74$~keV energy range with the (left) unshifted and (right) shifted ($+1.88$~eV) fayalite model in {\tt SolidFeL} (solid lines). These panels highlight the evidence for an energy offset error in the published experimental cross section data.  Similar results are found for the other species listed in Table~\ref{tab:compounds}.}
\end{figure}

Having established a well-defined continuum model, we then fit for the Fe-bearing compounds by replacing the silicate component of the {\tt ISMdust} model with one of the trial {\tt SolidFeL} compounds in Table~\ref{tab:compounds}. First, we refit the baseline continuum model across the full $0.45-1.80$~keV band, and then we restricted the fit to the Fe $L$-shell features around the narrow $0.69-0.74$~keV range, allowing only the abundance of the trial compound to vary. 

These {\tt SolidFeL} spectral fits yielded an unexpected and critical issue for their analysis, as the residuals displayed artifacts suggesting that the interstellar Fe absorption features are offset in energy compared to the laboratory measurements. An example of this is shown in Figure~\ref{fig:fayalite} for {\tt SolidFeL} fits for fayalite using both unshifted and shifted laboratory data. Similar artifacts were observed with every compound tried and for both sight lines. A notable feature of this comparison is the apparent emission at energies below the $L_3$ feature, which is solely an artifact of the inaccurate laboratory energy scale. 

Motivated by these artifacts, we reviewed the published literature reporting absolute energies for Fe $L$-shell XAFS. This is discussed in detail in Section~\ref{subsec:SolidPhaseFe} and summarized in Table~\ref{tab:L3}. There we found variances of $\sim 1-2$~eV among the published $L_3$ peak energies.  In principle, such a shift could also depend on the photon energy \citep{Muller:2021:PRA}; however, over the narrow energy range of the Fe $L$-shell features, such an effect is only a small correction to the overall shift and, thus, can be neglected here.

\begin{figure}[t]
{
\hfill
\centerline{\includegraphics[width=0.5\textwidth, clip=true]{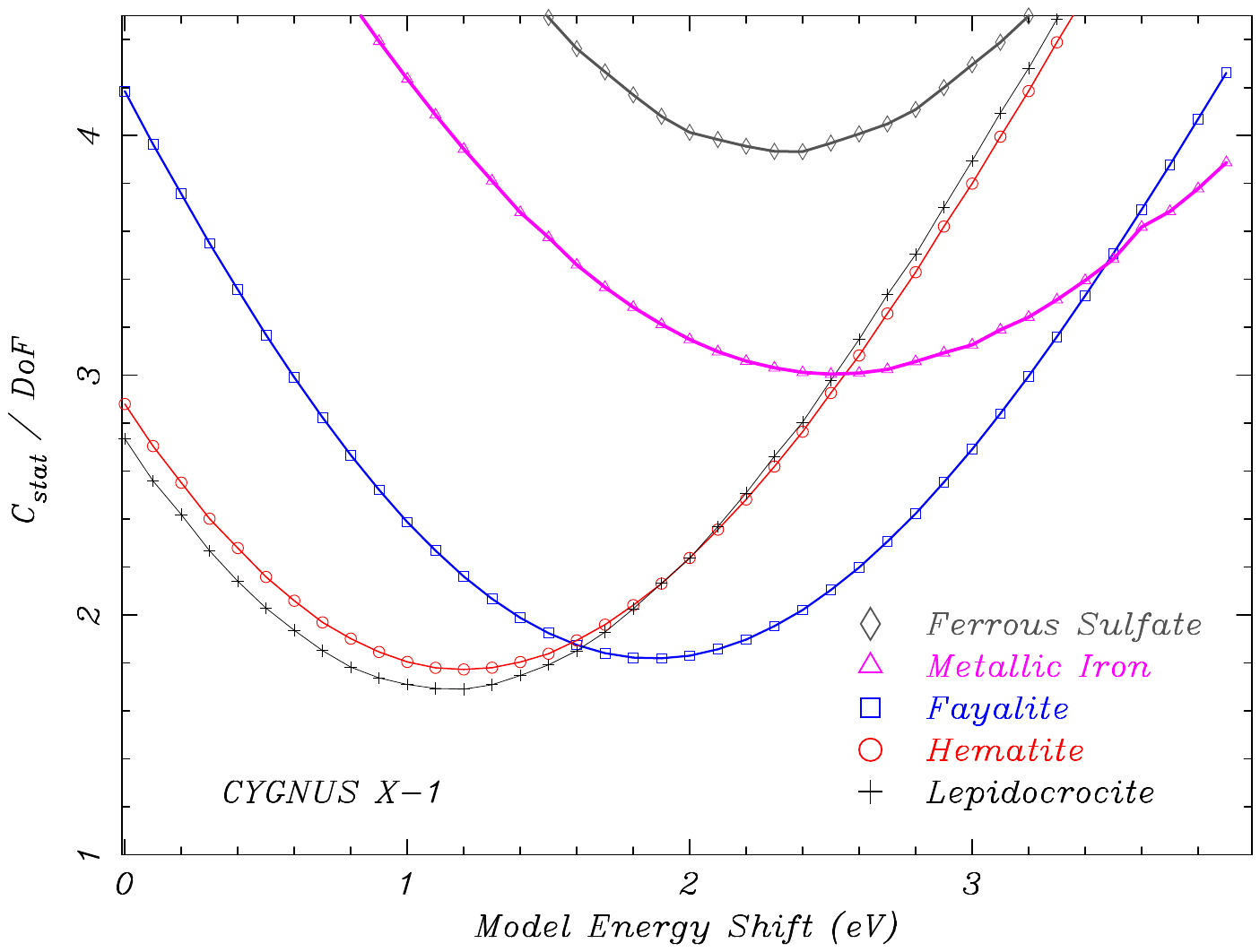}
\hfill
\includegraphics[width=0.5\textwidth, clip=true]{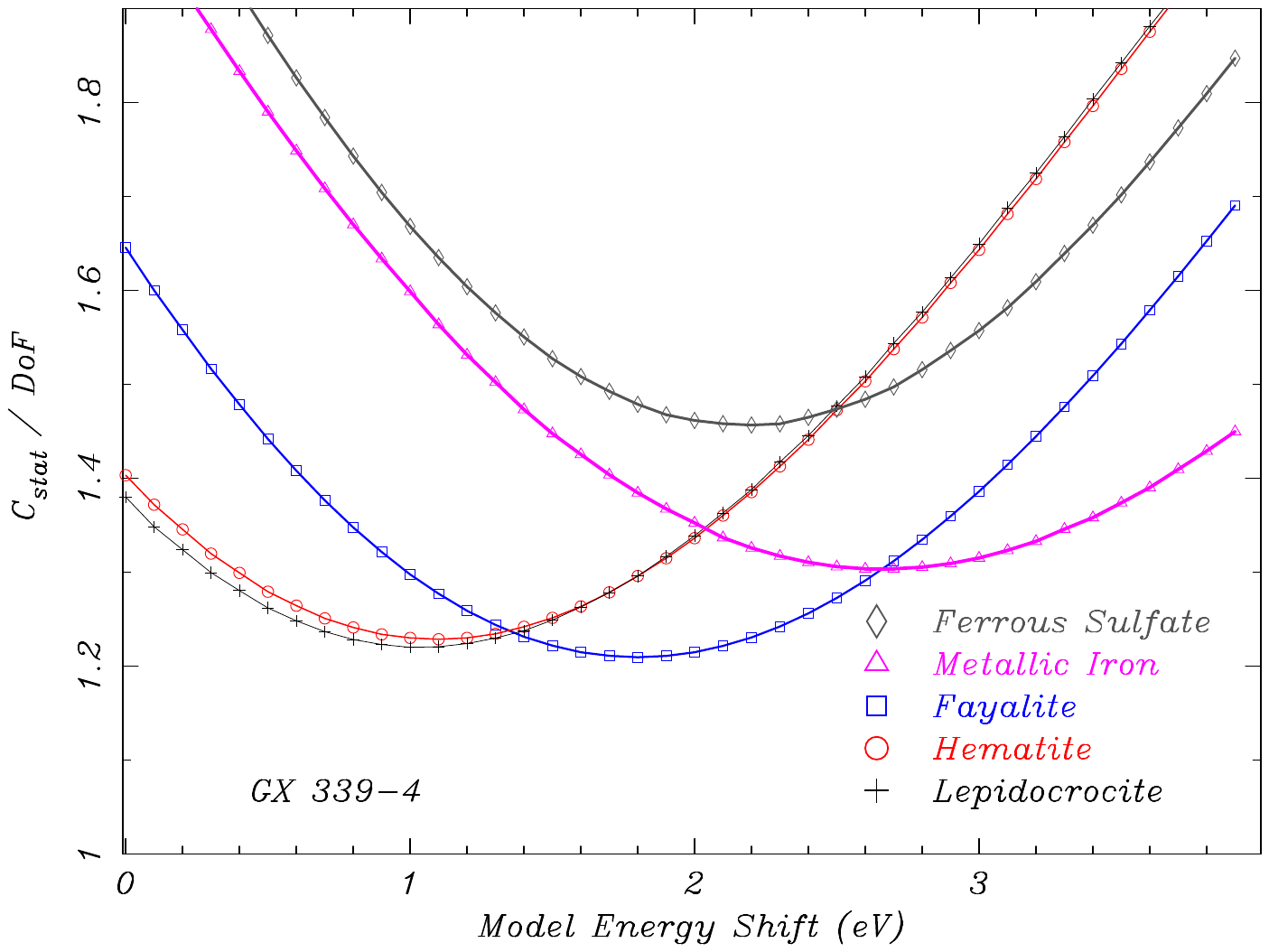}
\hfill
}
\vspace{-30pt}
\caption{\label{fig:ShiftedFits}
Calibration of the apparent offset error in the energy scale of the measured photoabsorption cross sections for the Fe-bearing compounds used in this work (see text for details). 
Plotted is the $C_{\rm stat}$ divided by the DoF for fits to the spectra of \cygnusx1\ (left)  and \gx339\ (right) as a function of an energy shift in 0.1~eV increments, for each {{\tt SolidFeL}} compound. These fits are restricted to the $0.69-0.74$~keV bandpass around the Fe $L$-shell features with the fitted $0.45 - 1.8$~keV continuum model held fixed. Note the change of the ordinate scale between the two panels.}
}
\end{figure}

To allow for an offset in the laboratory measurements, we introduced an energy shift to the photoabsorption data as a free parameter in our fits. Our new fitting method follows all of the procedures above, after systematically shifting the energy scale for each laboratory template in 0.1~eV increments for each of the five Fe-bearing compounds considered here (see Figure~\ref{fig:ShiftedFits}). Because the $L$-shell features are generally asymmetric and double-peaked, this method provided a reasonably accurate determination of the best fit for each trial compound using the shape of the extinction features.

\begin{figure}[t]
{
\hfill
\centerline{\includegraphics[width=0.5\textwidth, clip=true]{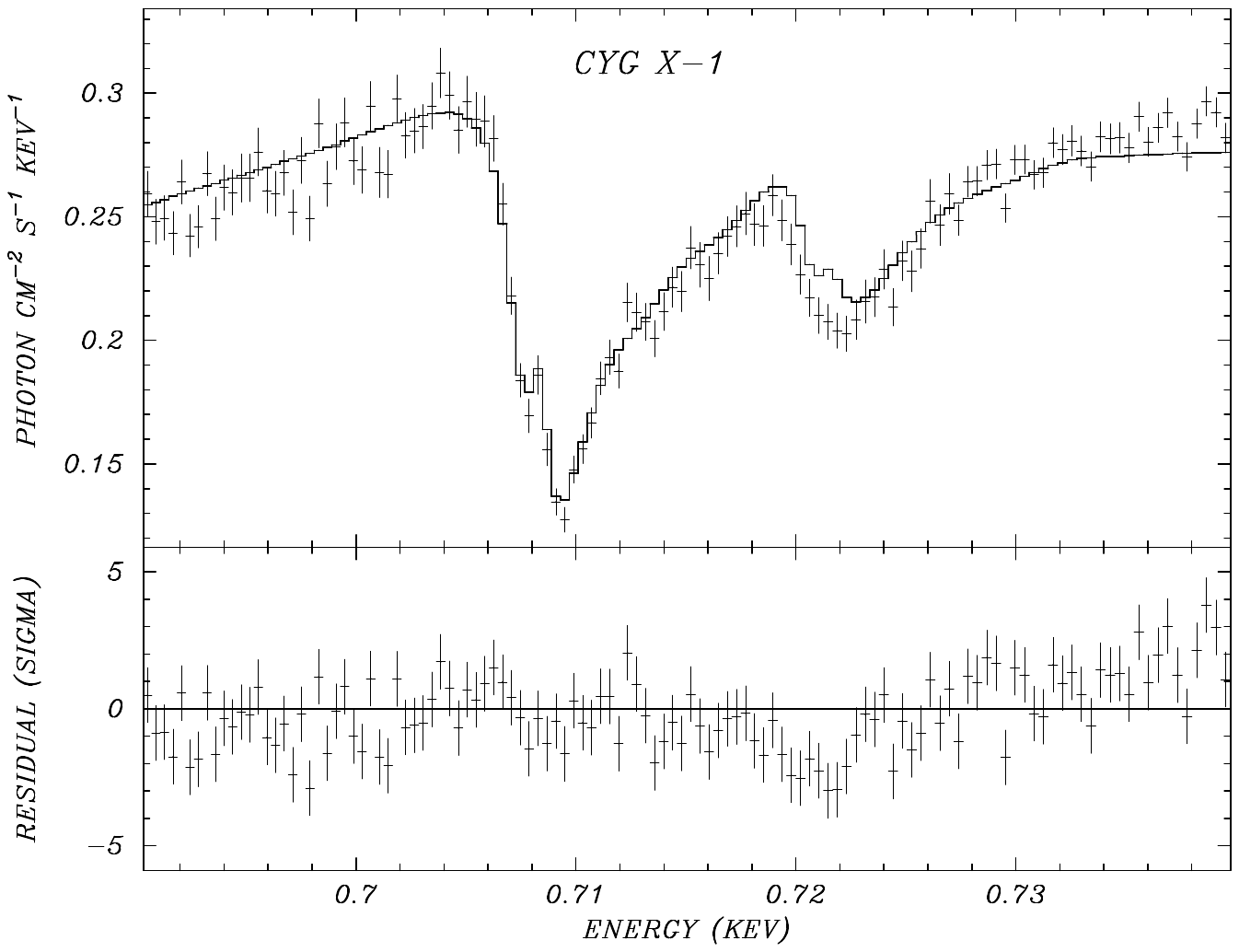}
\hfill
\includegraphics[width=0.5\textwidth, clip=true]{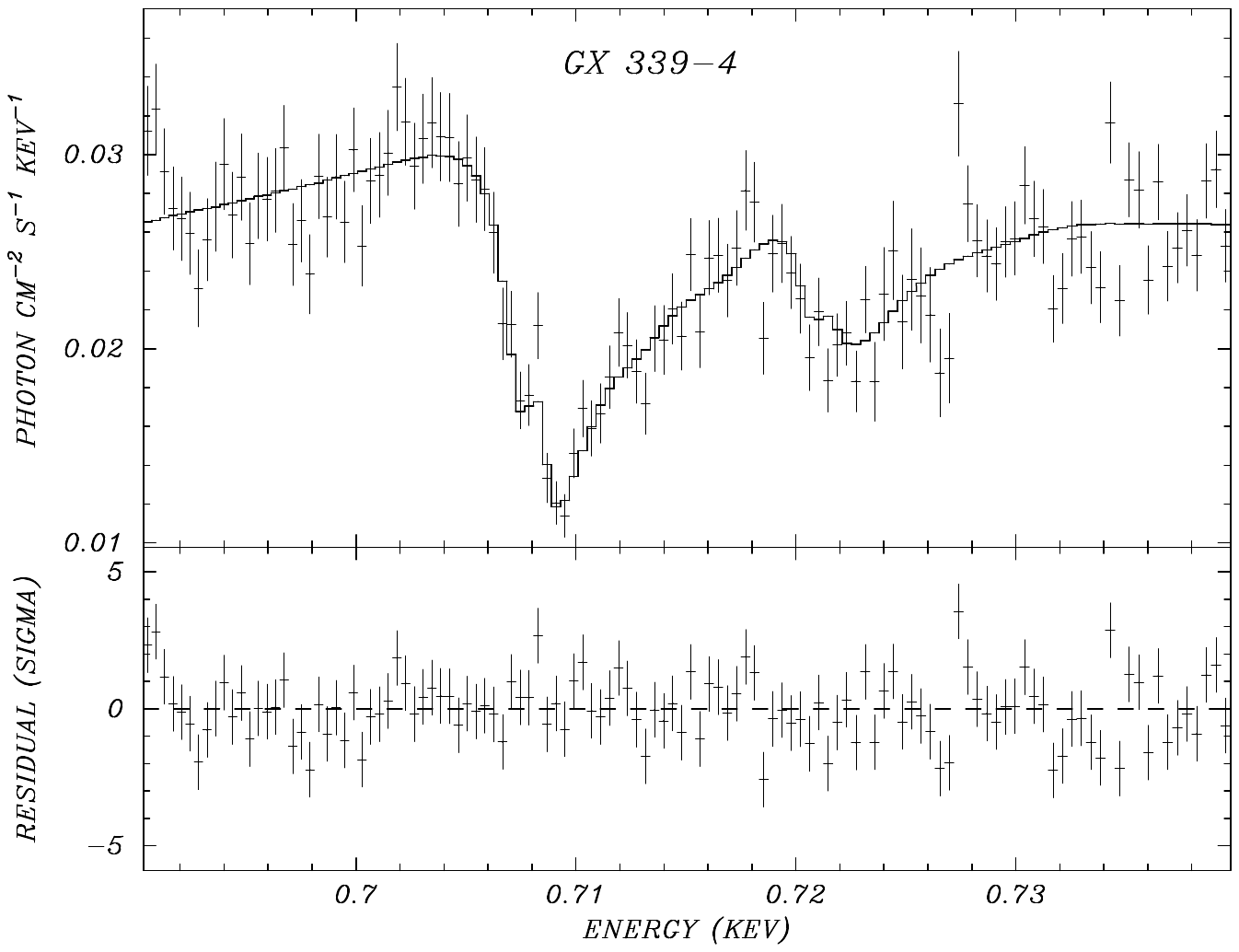}
\hfill
}}
\vspace{-30pt}
\caption{\label{fig:hematite}
RGS spectra highlighting the Fe $L$-shell extinction features of the diffuse ISM in the $0.69-0.74$~keV energy range for (left) \cygnusx1\ and (right) \gx339. The solid line shows the best-fit model for energy-shifted hematite.}
\end{figure}

The spectral fits that result from using the optimal shift is presented in Table~\ref{tab:bestfits} and the best-fit spectra are shown explicitly for the case of hematite in Figure~\ref{fig:hematite} for \cygnusx1\ and \gx339. The fits using energy-shifted lepidocrocite and fayalite for each sight line are nearly identical to those shown here (see, e.g., the right panel of Figure~\ref{fig:fayalite} for a fit to \cygnusx1 with energy-shifted fayalite). Table~\ref{tab:bestfits} demonstrates that, for a given compound, the results for both lines of sight imply the same energy shift and ranking of the goodness-of-fit statistic for that compound, which demonstrates that the need to shift the laboratory data for Fe $L$-shell absorption is not due to an anomaly in the Cygnus X-1 spectrum alone, as implied by previous work \citep[e.g.,][]{Juett:2006:ApJ, Hanke:2009:ApJ, Gatuzz:2015:ApJ}. 
For this study, we do not fit for multiple dust species at once because the uncertainty of the correct systematic shift for each laboratory measurement provides a free parameter space that is too large to draw meaningful conclusions at this time.

Lastly, we also examined the fit when gas-phase Fe~{\sc ii} was included.  The resulting Fe~{\sc ii} column density was consistent with zero with a three-sigma upper limit of 5\% of the total Fe column density for each compound. This is in agreement with past UV and optical studies that have found that over 90\% of interstellar Fe is in the solid phase. In the following, we present our conclusions from fits to the Fe $L$-shell photoabsorption features that include solid Fe-bearing compounds only.

\begin{deluxetable*}{|cc|ccc|ccc|}
\tablecaption{Narrow-band spectral fits to the Fe $L$-shell Features in the Two ISM Sight Lines Probed.\label{tab:bestfits}}
\tablewidth{0pt}
\tablehead{
                  &           &  \multicolumn{3}{c|}{Cygnus~X-1}                & \multicolumn{3}{c|}{GX~339-4}\\
{{\tt SolidFeL}}  &           & $\Delta E_{\rm shift}$ & $M_{{\rm d},i}$ & $C_{\rm stat}$   & $\Delta E_{\rm shift}$ & $M_{{\rm d},i}$ & $C_{\rm stat}$\\
       {Compound $i$} & {Formula} & (eV)               & ($10^{-4}~{\rm g~cm}^{-2}$)  &    (${\rm DoF} = 119$)        &  (eV) &\colhead{($10^{-4}~{\rm g~cm}^{-2}$)} &   (${\rm DoF} = 119$)
}
\startdata
    Lepidocrocite   & $\gamma$-FeOOH       & $1.14\pm0.26$ & $0.301\pm0.022$ & 201.1& $1.04\pm0.61$ & $0.349\pm0.047$ & 145.1\\ 
    Hematite        & $\alpha$-Fe$_2$O$_3$ & $1.21\pm0.26$ & $0.320\pm0.022$ & 211.0& $1.10\pm0.61$ & $0.376\pm0.050$ & 145.2\\ 
    Fayalite        & Fe$_2$SiO$_4$        & $1.88\pm0.27$ & $0.277\pm0.019$ & 216.3& $1.80\pm0.63$ & $0.318\pm0.043$ & 143.9\\ 
    Ferrous Sulfate & FeSO$_4$             & $2.32\pm0.26$ & $0.394\pm0.032$ & 467.7& $2.18\pm0.59$ & $0.533\pm0.082$ & 173.4\\ %
    Metallic Iron   & Fe                   & $2.51\pm0.31$ & $0.252\pm0.009$ & 357.6& $2.66\pm0.71$ & $0.290\pm0.026$ & 155.1 
\enddata
\tablecomments{The calibrated energy shift $\Delta E_{\rm shift}$ is applied in the laboratory frame of the cross sections for each compound (see text). The mass column is from the fits to the Fe features in the $0.69 - 0.74$~keV range using the {{\tt SolidFeL}} model with the fitted $0.45 - 1.8$~keV continuum model held fixed. Uncertainties are quoted at a 90\% confidence level for one interesting parameter.}
\end{deluxetable*}

\section{Discussion}
\label{sec:Discussion}

Given the $1-2$~eV spread in the published laboratory data for the Fe~$L_3$ photoabsorption feature, it is difficult to constrain the mineralogy of the interstellar dust along the two sight lines based solely on the best-fit energy shifts reported in Table~\ref{tab:bestfits}. However, a few conclusions can be made based on the shape of the Fe L photoabsorption features relative to the laboratory results. First, ferrous sulfate produces the worst fit for each sight line, leading us to conclude that this compound is unlikely to be abundant in the diffuse ISM. The second worst fitting compound is that of an MRN distribution of metallic Fe particles, no matter what energy shift is applied. 
This leads us to conclude that metallic Fe is not likely to be the dominant carrier of Fe in the diffuse ISM. Our study, however, does not necessarily rule out the presence of metallic Fe nanoparticles, which we discuss in Section~\ref{subsec:Astrophysics}.

This leaves lepidocrocite, hematite, and fayalite as the three most viable interstellar Fe-bearing compounds in the diffuse ISM, for the sight lines and analyses presented here.  However, it is difficult to determine the most likely of the three as their best-fit statistics are so similar.  In addition, the energy shifts for lepidocrocite and hematite are nearly identical.  The similarity in the fits for these two species is because their photoabsorption structures are so alike in shape and magnitude that current X-ray astronomy spectrometers cannot distinguish the subtle differences between the two compounds.  We might, however, be able to disfavor fayalite, in part because of abundance constraints as we discuss later.

\subsection{Abundances}
\label{subsec:Abundances}

The best-fit model free parameters, using the optimal energy shift for each Fe-bearing compound, are given in Tables~\ref{tab:cygx1dust} and \ref{tab:gx339dust}. The tables also report the corresponding O, Fe, and Ne abundances per H atom for each of these fits, reported as $\log A_Z + 12$, where $A_Z$ is the number of atoms per H atom for the element $Z$.  To convert the mass column $M_{{\rm d},i}$ for each Fe-bearing species $i$ into an Fe atom column density, we multiply by the value of $N_{\rm Fe}$ given in Table~\ref{tab:compounds}. We found that the percent error on the free parameters listed in both tables are identical to those in Table~\ref{tab:specresults}. We use those percent errors in reporting the propagated error on the abundance measurements in Tables~\ref{tab:cygx1dust} and \ref{tab:gx339dust}. 
Below, we compare these abundances to Solar abundances \citep{Asplund:2021:AA} and to the canonical ISM abundances used in X-ray spectroscopy \citep{Wilms:2000:ApJ}.

Note that the Mg abundances reported in Tables~\ref{tab:cygx1dust} and \ref{tab:gx339dust} are non-physical because we replaced the \texttt{ISMdust} silicate model with the \citet{Lee:2009:ApJ} Fe-bearing compounds, none of which contained Mg. This causes the best gas-phase Mg component from \texttt{ISMabs} to get larger, compared to the results shown in Table~\ref{tab:specresults}, in order to make up for the missing extinction component that was present in the \texttt{ISMdust} silicate model. Furthermore, there is no specific stoichiometry associated with the \citet{Draine:2003:ApJ} silicate opacities used in \texttt{ISMdust}, so a physically relevant estimation of the Mg column from $M_{\rm sil}$ in Table~{\ref{tab:specresults} is not possible in this work.

\begin{deluxetable*}{cccccccc}
\tablecaption{Best fit results for \cygnusx1\ with dust model energy scale shifted \label{tab:cygx1dust}}
\tablewidth{0pt}
\tablehead{
\colhead{Model}  & \colhead{Parameter} & \colhead{Unit}  & \colhead{Fayalite} & \colhead{Hematite} & 
\colhead{Lepidocrocite} & \colhead{Metallic Fe}      & \colhead{Ferrous Sulfate} }
\startdata
{\tt ISMabs} & $N$(H) & 10$^{22}$~cm$^{-2}$ & 0.582 & 0.596 & 0.592 & 0.592 & 0.574 \\
& $N$(O~{\sc i}) & 10$^{16}$cm$^{-2}$ & 366 & 371 & 363 & 383 & 354 \\
& $N$(O~{\sc ii}) & 10$^{16}$cm$^{-2}$ & 1.1 & 1.4 & 1.1 & 1.7 & 0.6 \\
& $N$(O~{\sc iii}) & 10$^{16}$cm$^{-2}$ & 10 & 10 & 9 & 12 & 8 \\
& $N$(Ne~{\sc i}) & 10$^{16}$cm$^{-2}$ & 57 & 56 & 54 & 59 & 54 \\
& $N$(Mg~{\sc ii}) & 10$^{16}$cm$^{-2}$ & 32.0 & 32.5 & 32.5 & 31.6 & 32.6 \\
\hline
Power-law & $\mathit{\Gamma}$ & unity & 2.27 & 2.34 & 2.30 & 2.25 & 2.35 \\
& Norm & arbitrary & 2.13 & 2.30 & 2.20 & 2.18 & 2.26 \\
\hline
Gaussian & $E_1$ & keV & 0.832 & 0.832 & 0.831 & 0.834 & 0.830 \\
& Norm$_1$ & arbitrary & 0.059 & 0.066 & 0.067 & 0.056 & 0.073 \\
& $E_2$ & keV & 0.681 & 0.678 & 0.681 & 0.700 & 0.668 \\
& Norm$_2$ & arbitrary & 0.078 & 0.064 & 0.072 & 0.055 & 0.106 \\
& $\sigma_{1,2}$ & keV & 0.036 & 0.036 & 0.036 & 0.036 & 0.036 \\
\hline
{\tt SolidFeL} & $M_{{\rm d},i}$ & 10$^{-4}$ g cm$^{-2}$ & $0.277 \pm 0.019$ & $0.320 \pm 0.022$ & $ 0.301 \pm 0.022$ & $0.252 \pm 0.009$ & $0.394 \pm 0.032$  \\
\hline
$C_{\rm stat}$ & &  & 216.3 & 211.0 & 201.1 & 357.6 & 467.7 \\
\hline
\multicolumn{3}{c}{$\log A_{\rm O (gas)} + 12$} & $8.813\pm0.009$ & $8.805\pm0.009$ & $8.802\pm0.009$ & $8.828\pm0.009$ & $8.804\pm0.008$ \\
\multicolumn{3}{c}{$\log A_{\rm O (Fe-dust)} + 12$} & $7.752\pm0.030$ & $7.781\pm0.030$ & $7.840\pm0.032$ & -- & $8.040\pm0.035$ \\
\multicolumn{3}{c}{$\log A_{\rm Fe} + 12$} & $7.451\pm0.030$ & $7.605\pm0.030$ & $7.539\pm0.032$ & $7.663\pm0.016$ & $7.438\pm0.035$ \\
\multicolumn{3}{c}{$\log A_{\rm Ne} + 12$} & $7.995\pm0.033$ & $7.974\pm0.033$ & $7.964\pm0.033$ & $8.001\pm0.033$ & $7.975\pm0.033$ \\
\enddata
\tablecomments{Only free parameters are listed. When not listed, the corresponding uncertainties are given in Tab.~\ref{tab:specresults}. The DOF is 119.}
\end{deluxetable*}

From our solid-phase Fe fits, we obtain an Fe/H abundance that is close to solar \citep[$\log A_{\rm Fe} + 12 = 7.46 \pm 0.04$,][]{Asplund:2021:AA} and ISM \citep[$\log A_{\rm Fe} + 12 = 7.43$,][]{Wilms:2000:ApJ}, but generally larger by $\sim 0.1-0.3$~dex ($25-100\%$). When we included gas-phase Fe ions in our modeling, we obtained gas-phase Fe abundances that were consistent with zero. Thus we are unable to constrain the abundance of gas-phase Fe with these X-ray data alone.  Given that $90-99\%$ of the Fe is locked up in dust, accounting for the missing gas-phase Fe will only increase the total cold-phase Fe abundance for both sight lines by  $\sim 0.004-0.046$~dex. This is on the order of the abundance measurement uncertainties. 

Of the top three compounds that best fit the diffuse ISM data, fayalite puts the total Fe content closest to solar, while hematite and lepidocrocite suggest over-abundances on the $\sim 5\sigma$ level. Along the Cygnus X-1 sight line, the Fe abundance is consistent with solar, if fayalite is the dominant carrier of Fe in the diffuse ISM; whereas for GX~339-4, the Fe abundance is $\sim 30\%$ larger. If the iron oxides hematite and lepidocrocite account for the majority of Fe-bearing interstellar dust, then the Fe/H abundance toward Cygnus~X-1  would be a factor of $\sim 1.4$ higher than solar and toward GX~339-4 a factor of $\sim 2$ higher.

\begin{deluxetable*}{cccccccc}
\tablecaption{Best fit results for GX339--4 with dust model energy scale shifted \label{tab:gx339dust}}
\tablewidth{0pt}
\tablehead{
\colhead{Model}  & \colhead{Parameter} & \colhead{Units}  & \colhead{Fayalite} & \colhead{Hematite} & 
\colhead{Lepidocrocite} & \colhead{Metallic Fe}      & \colhead{Ferrous Sulfate} }
\startdata
{\tt ISMabs} & $N$(H) & 10$^{22}$~cm$^{-2}$ & 0.51 & 0.53 & 0.52 & 0.52 & 0.50 \\
& $N$(O~{\sc i}) & 10$^{16}$~cm$^{-2}$ & 316 & 323 & 314 & 335 & 299 \\
& $N$(O~{\sc ii}) & 10$^{16}$~cm$^{-2}$ & 6.3 & 7.3 & 6.1 & 8.6 & 4.8 \\
& $N$(O~{\sc iii}) & 10$^{16}$~cm$^{-2}$ & 9 & 10 & 8 & 12 & 7 \\
& $N$(Ne~{\sc i}) & 10$^{16}$~cm$^{-2}$ & 49 & 46 & 45 & 53 & 38 \\
& $N$(Mg~{\sc ii}) & 10$^{16}$~cm$^{-2}$ & 29 & 30 & 30 & 29 & 31 \\
\hline
Power-law & $\mathit{\Gamma}$ & unity & 2.62 & 2.70 & 2.67 & 2.62 & 2.77 \\
& Norm & arbitrary & 0.149 & 0.164 & 0.155 & 0.154 & 0.169 \\
\hline
Gaussian & $E_1$ & keV & 0.824 & 0.822 & 0.821 & 0.826 & 0.818 \\
& Norm$_1$ & arbitrary & 0.0018 & 0.0023 & 0.0024 & 0.0015 & 0.0035 \\
& $E_2$ & keV & 0.672 & 0.663 & 0.670 & 0.700 & 0.661 \\
& Norm$_2$ & abritrary & 0.0039 & 0.0030 & 0.0034 & 0.0014 & 0.0068 \\
& $\sigma_{1,2}$ & keV & 0.032 & 0.032 & 0.032 & 0.032 & 0.032 \\
\hline
{\tt SolidFeL} & $M_{{\rm d},i}$ & 10$^{-4}$ g cm$^{-2}$ & $0.318 \pm 0.043$ & $0.376 \pm 0.050$ & $0.349 \pm 0.047$ & $0.290 \pm 0.026$ & $0.533 \pm 0.082$ \\
\hline
$C_{\rm stat}$ & &  & 143.9 & 145.2 & 145.1 & 155.1 & 173.4 \\
\hline
\multicolumn{3}{c}{$\log A_{\rm O (gas)} + 12$} & $8.813\pm0.009$ & $8.807\pm0.010$ & $8.800\pm0.009$ & $8.834\pm0.010$ & $8.793\pm0.009$ \\
\multicolumn{3}{c}{$\log A_{\rm O (Fe-dust)} + 12$} & $7.868\pm0.059$ & $7.905\pm0.058$ & $7.959\pm0.059$ & -- & $8.228\pm0.065$ \\
\multicolumn{3}{c}{$\log A_{\rm Fe} + 12$} & $7.566\pm0.059$ & $7.728\pm0.058$ & $7.658\pm0.059$ & $7.779\pm0.039$ & $7.626\pm0.065$ \\
\multicolumn{3}{c}{$\log A_{\rm Ne} + 12$} & $7.984\pm0.033$ & $7.938\pm0.033$ & $7.936\pm0.033$ & $8.006\pm0.033$ & $7.886\pm0.033$ \\
\enddata
\tablecomments{Only free parameters are listed. When not listed, the corresponding uncertainties are given in Tab.~\ref{tab:specresults}. The DOF is 119.}
\end{deluxetable*}

Using these results, we can determine Fe-bearing dust-to-hydrogen mass ratio for dust-grain compound $i$ from
\begin{equation}
     X_{{\rm d},i/{\rm H}} = \frac{M_{{\rm d},i}}{\NH m_{\rm p}},
\end{equation}
where $m_{\rm p}$ is the proton mass. Assuming the average dust-to-gas mass ratio of the Milky Way ISM is 0.01 \citep{Draine:2011:Book}, then $X_{\rm d/H}$ for all dust compounds should sum to 0.014, for a He/H number ratio of 10\%. The average mass ratio of Fe-bearing dust grains from our best fitting candidates are $X_{\rm Fe-dust/H} = 0.003$ and $0.004$ for Cygnus~X-1 and GX~339-4, respectively. This suggests that Fe-bearing grains make up $\sim 20-30\%$ of the dust-to-hydrogen mass expected in the ISM. 

The total abundance of cool gas-phase O is determined here from the {\tt ISMabs} fits to the three O charge states detected. For both sight lines, we find a gas-phase O abundance that is larger by $\sim 0.1$~dex ($\sim 30\%$) than the solar \citep[$\log A_{\rm O} + 12 = 8.69 \pm 0.04$,][]{Asplund:2021:AA} and ISM \citep[$\log A_{\rm Fe} + 12 = 8.69$,][]{Wilms:2000:ApJ} values. 

We can also infer the abundance of O locked up in dust grains. During the fit procedure, we replaced the \texttt{ISMdust} silicate model with {\tt SolidFeL}, which includes only the laboratory measurements for Fe-bearing compounds, a few of which contained O, but no Mg-bearing silicates.  Thus we are only able to consider O abundances from Fe-bearing dust in this work, reported as $A_{\rm O(Fe-dust)} + 12$ in Tables~\ref{tab:cygx1dust} and \ref{tab:gx339dust}. For our top three dust candidates, the abundance of O in Fe-bearing grains is about 1~dex lower than in the gas phase.

Mg-rich silicates are suspected to account for the majority of ISM silicates and to be an ample reservoir for solid-phase O. 
Assuming that 100\% of interstellar Si is depleted into the Mg-silicate end-members enstatite (MgSiO$_3$) or forsterite (Mg$_2$SiO$_4$), then $\log A_{\rm O(Mg-sil)} + 12$ would be $7.75$ and $7.87$, respectively, using the ISM abundances of \citet{Wilms:2000:ApJ}. 
These values are comparable to our reported $A_{\rm O(Fe-dust)}$ values. This implies that if Fe oxides are indeed the dominant Fe-bearing dust compound, then the amount of ISM O depleted into Fe oxides is comparable to that depleted into Mg silicates.  Adding together the inferred abundances for $A_{\rm O(Fe-dust)}$ and $A_{\rm O(Mg-sil)}$, the Fe-oxide models imply a total O depletion $\sim 20\%$. Depletions of $10-30\%$ are reasonable for the diffuse ISM \citep{Jenkins:2009:ApJ, Psaradaki:2020:AA, Psaradaki:2023:AA}. 
If we assume that the majority of interstellar silicates are in enstatite, a requirement due to the fact that the abundance of Si is approximately equal to that of Mg, then we can estimate the total mass of Mg-rich silicate dust based on the mass and mixture of each element in the compound stoichiometry and the abundance of Si relative to H. This yields $X_{\rm Mg-sil/H} \approx 0.002$, which accounts for $\sim 10\%$ of the interstellar dust mass.  
Together with $X_{\rm Fe-dust/H} = 0.003-0.004$, this would leave the next suspected dust species, carbonaceous grains, to account for the remaining  $\sim 60\%$ of the total interstellar dust mass.

Since the abundances of Mg, Fe, and Si are close to one another, assuming that fayalite (our third-best fitting model) is the dominant Fe-bearing dust compound would necessarily reduce the amount of Si available to form enstatite or forsterite. The fayalite fit suggests solid-Fe abundances that are close to Solar (i.e., accounting for all known solid-phase Fe), implying that the majority of Si must also be in fayalite. This is inconsistent with the current understanding of interstellar silicate mineralogy, for which Mg/Fe ratios $\approx 1-3$ are found (see Sections~\ref{sec:Iron} and \ref{subsec:Comparison}). With little Si remaining in the budget to form Mg-rich silicates, the fayalite fit yields an O depletion $\sim 10\%$.  The fayalite model also yields $X_{\rm Mg-sil/H} \approx 0.003-0.004$, for Cygnus~X-1 and GX~339--4, respectively, implying that $20-30\%$ of interstellar dust is in Fe-rich silicates. However, with relatively little Si remaining to form Mg-rich silicates, the remaining $70-80\%$ are presumably in carbonaceous grains. This picture is broadly inconsistent with the majority of mineralogical models of interstellar dust.

Since Ne is a noble gas that does not deplete into the solid phase, it provides an excellent benchmark for abundance constraints. 
Our measured values of $\log A_{\rm Ne} +12 = 7.94 - 8.00$ are slightly underabundant by $0.05 - 0.1$~dex ($10-30\%$) compared to studies of the local solar neighborhood \citep[$8.09 \pm 0.05$,][]{Nieva:2012:AA}, the Orion Nebula \citep[$0.05 \pm 0.03$][]{SimonDiaz:2011:AA}, and the solar wind \citep[$8.06 \pm 0.05$][]{Asplund:2021:AA}. Curiously, our measurements are in better agreement with less modern values for the Ne abundance of the ISM \citep[$7.94$,][and references therein]{Wilms:2000:ApJ}. 
Note that, if we were to adopt a lower value for  $N({\rm H})$ based on the observed Ne column density, it would cause the relative over-abundances of O and Fe reported above to be even more pronounced. 
However, our observed Ne abundances are within 2$\sigma$ of the expected values, which is not significant enough to challenge our current assumptions.

Overall, there are major systematic difficulties in determining the total ISM abundance of H from X-ray spectroscopy. That is because H has no spectrocopic signature in the $0.3-10$~keV band and only contributes to continuum absorption. 
Thus, the $N({\rm H})$ value measured from the X-ray band is inferred indirectly by continuum fitting from an assumed abundance table and, in this work, an assumed dust-to-gas mass ratio of various compositions. In the $< 0.5$~keV super-soft X-ray band, continuum absorption by C is strong, and instruments are at the limits of sensitivity. As a result, the spectroscopic signature of C is rarely measurable. Our X-ray spectra are thereby unable to constrain the C/H abundance or C-bearing dust mass column directly, and the inferred H abundance will change depending on those assumptions. The estimated abundance of carbonaceous dust relative to silicates varies widely from approximately equal \citep{Draine:2003:ApJ, Corrales:2016:MNRAS} to grain models that contain almost no carbonaceous dust except for PAHs \citep{Zubko:2004:ApJS, Hensley:2023:ApJ}.  
Here, we have tied the C abundance to the H abundance following the assumptions present in \citet{Corrales:2016:MNRAS}. However, since C is a significant source of soft X-ray extinction, the true mass ratio of carbonaceous grains relative to H is a source of systematic uncertainty for all abundance determinations in the X-ray.

\subsection{Astrophysical Implications}
\label{subsec:Astrophysics}

The ability to predict the mineralogical form of interstellar Fe from Solar System analogs is limited, due to refractory processing by the solar protoplanetary disk. Nonetheless, these analogs provide a good starting point for the materials that can form in the diffuse ISM. 
Suspected presolar grains and cometary dust collected from the comet Wild 2 contain Fe in the form of FeS and Fe-Ni metal with $<2 \%$ Ni by number \citep{Bradley:1994:Science, Zolensky:2006:Science, Leroux:2008a:MPS, Leroux:2008b:MPS}
The closest analog present in our study is ferrous sulfate (FeSO$_4$), which is the worst fitting compound for both sight lines. Other studies have also not found substantial signatures of iron sulfides in the ISM (see Section~\ref{subsec:Comparison}).  Our findings rule out ferrous sulfate as a major reservoir of interstellar Fe. 

A number of studies have proposed metallic Fe as the main carrier of interstellar Fe, from both a theoretical and observational perspective. 
Pure metallic iron nanoparticles, either embedded in silicates or later exposed as the silicates erode in the warmer phases of the ISM, are proposed to explain the observed Fe-depletion patterns of the ISM \citep{Zhukovska:2018:ApJ}. Continuum extinction from pure metallic iron is also invoked in order to explain the full spectrum of absorption in dusty circumstellar outflows \citep{Harwit:2001:ApJ, Kemper:2002:Nature}. 
However, recent laboratory experiments failed to produce pure metallic Fe particles in a space-like environment \citep{Kimura:2017:SciAdv}. 
In our study here, metallic Fe with an MRN distribution provided the second to worst fit even when we allow for different energy scale calibrations.  
We conclude that the majority of interstellar Fe is not an MRN distribution of grains in metallic form, though metallic Fe nanoparticles may still account for a small fraction of the total ISM Fe abundance.  

From both a theoretical and observational standpoint, interstellar silicates are expected to be Mg-rich. Enstatite 
and forsterite, 
the Mg-rich end-members of olivines (Mg$_{2x}$Fe$_{2(1-x)}$SiO$_4$) and pyroxenes (Mg$_x$Fe$_{(1-x)}$SiO$_3$), respectively, condense at higher temperatures than Fe-enriched silicates \citep{Gail:2010:Springer}. Thus for silicates formed in AGB stellar outflows, the Mg-rich silicates likely freeze out before they can be enriched in Fe. 
Collected samples of suspected presolar grains and cometary dust also contain silicates almost entirely comprised of Mg end-members 
\citep{Bradley:1994:Science, Bradley:2010:Springer, Hanner:2010:Springer, Mann:2010:Springer}. 
As described in Section~\ref{subsec:Abundances}, a dust model where fayalite accounts for the majority of solid interstellar Fe is physically inconsistent with the low abundance of Fe-bearing silicates measured both in situ and from the IR signatures in the ISM. Fitting fayalite to the observed Fe $L$-shell photoabsorption features also requires a shift in energy scale calibration that is quite large compared to the majority of laboratory measurements (Table~\ref{tab:L3}). For these reasons, fayalite is disfavored by our analysis as the predominant form of Fe-bearing dust.

Chemically active O is cosmically the most abundant element heavier than chemically inert He, making it highly likely that gas-phase Fe will bind with O in the ISM. This warrants the exploration of iron oxides, such as hematite and lepidocrocite, here, despite both of them being absent from metereotic and interplanetary dust particles. Unfortunately, the extinction features of these two compounds are nearly identical when the throughput and spectral resolution of modern day X-ray telescopes and spectrometers are considered; the compounds cannot be distinguished from one another. 
Nonetheless, these are the two best-fitting compounds that require the smallest energy shift with respect to the published laboratory work. This leads us to hypothesize that Fe$^{3+}$-bearing compounds are most likely to be the dominant carriers of solid-phase Fe in the ISM. If this is the case, then interstellar Fe may be more abundant relative to solar than previously estimated. However, systematic uncertainties in measuring $\NH$ accurately in the X-ray band makes it difficult to 
draw definitive conclusions about iron oxides via abundance arguments.

If iron oxides truly are the most abundant component of solid-phase Fe in the ISM, then they provide a distinct reservoir for interstellar O that has not been highly explored. 
As described in Section~\ref{subsec:Abundances}, the amount of O depleted into iron oxides could be comparable to that depleted into Mg silicates. 
This is of interest for the oxygen problem -- namely, that the observed O depletion rate in the ISM cannot be explained entirely by its incorporation by Si, Mg, and Fe compounds \citep{Jenkins:2009:ApJ}. 
However, the sight lines studied here are unlikely to resolve this issue. 
For a distance of 2.2~kpc to Cygnus~X-1 \citep{Arnason:2021:MNRAS} and 5~kpc to GX~339-4 \citep[Gaia DR3,][]{Gaia:2020:Cat}, and using our derived $N({\rm H})$ values, the average density for each sight line is $\left< n_{\rm H} \right> \approx 0.9$ and $0.3$~cm$^{-3}$, respectively. One molecular cloud with a radius $\sim 100$~pc and $n_{\rm H} \sim 10^2$~cm$^{-3}$ in the sight line would add $\sim 3 \times 10^{22}$~cm$^{-2}$ to $N({\rm H})$, which is $\sim 5-6$ times higher than our observed $\NH$ values. 
Hence, our observations do not probe the state of interstellar dust particles in the densest regions of molecular clouds, where the oxygen problem may be solved by the formation of O-rich ices.

\subsection{Comparison to other work}
\label{subsec:Comparison}

Recent laboratory measurements of over a dozen compounds have enabled new X-ray astromineralogical studies of Fe-bearing grains through Fe extinction directly \citep{Rogantini:2018:AA, Westphal:2019:ApJ, Psaradaki:2023:AA} or indirectly through Si, Mg, and O absorption features  \citep{Zeegers:2017:AA, Zeegers:2019:AnA, Rogantini:2019:AA, Rogantini:2020:AnA, Psaradaki:2020:AA}. Those studies all used MRN distributions.  However, the photoabsorption cross section data from those studies are not yet publicly available. This prevented our being able to incorporate those results into {\tt SolidFeL} for abundance determinations and also prevented our being able to explore the calibration accuracy of the absolute energy scale for their cross section data.  Within these constraints, we discuss here their findings relative to ours.

Comparing the astronomical results based of the above new laboratory data, a unified, comprehensive picture of Milky Way interstellar dust has yet to emerge. Examination of Mg~$K$- and Si~$K$-shell features suggest $\sim 70-80\%$ of silicates by number are amorphous olivine with ${\rm Mg/Fe} \sim 1$ \citep[MgFeSiO$_4$;][]{Rogantini:2019:AA, Zeegers:2019:AnA, Rogantini:2020:AnA}. On the other hand, direct measurement of the Fe~$L$-shell extinction data for five XRBs suggests that the most prevalent silicates are amorphous, Mg-rich pyroxenes with ${\rm Mg/Fe} = 3$ (Mg$_{0.75}$Fe$_{0.25}$SiO$_3$), constituting by number about 70\% of the Fe- and O-bearing dust along the line of sight \citep{Psaradaki:2023:AA}. In that study, only the sight line of GX~9+9 was best fit with a dust column where the majority of Fe- and O-bearing dust was amorphous olivine with ${\rm Mg/Fe}= 1$, in agreement with the previous Mg~$K$- and Si~$K$-shell studies. 

Our {\tt SolidFeL} study included the olivine fayalite, but not a pyroxene compound.  However, both olivine and pyroxene are Fe$^{2+}$-bearing minerals and are likely to peak in their photoabsorption cross section at similar energies (as was seen for the Fe$^{3+}$ minerals lepidocrocite and hematite). Thus we suggest that fayalite serves as a reasonable proxy for pyroxene in this work.  Unfortunately, if Fe-rich silicates are indeed the majority host material for interstellar Fe, then there is insufficient interstellar Si available to provide significant dust mass in Mg-rich silicates (demonstrated in Section~\ref{subsec:Abundances}).  This picture is incongruent with the majority of astromineralogical studies, from both the X-ray (described above) and the IR \citep{Min:2007:AA}  perspectives. However, though this picture is generally disfavored in our analysis, systematic uncertainties in the $\NH$ column means that we cannot not completely exclude fayalite as an Fe-bearing constituent of the diffuse ISM based on these abundance arguments alone. 

\citet{Psaradaki:2023:AA} also conclude that $10-20\%$ of the dust column is made up of metallic iron, after shifting the  \citet{Kortright:2000:PhRvB} cross section by $-0.3$~eV to match the position observed by \citet{Fink:1985:PRB}, listed in Table~\ref{tab:L3}, whereas we needed to shift the metallic Fe data by $\approx +2.53$~eV. 
We note that 10\% is also on the order of magnitude expected for the abundance of gas-phase Fe in the diffuse ISM. Our fit procedures were unable to constrain the abundance of gas-phase Fe, given its small contribution to the absorption signal. It follows that we are likely unable to comment on a similar abundance of metallic Fe particles.
Nonetheless, the fact that the \citet{Kortright:2000:PhRvB} metallic Fe cross section requires shifting in nearly every study of interstellar Fe \citep[e.g.,][]{Juett:2006:ApJ, Hanke:2009:ApJ, Gatuzz:2015:ApJ, Psaradaki:2023:AA}, as well as in our work, demonstrates that the absolute energy calibration of pure metallic iron is essential for accurate identification and abundance determinations of interstellar dust, and needs to be revisited.

Additional compounds examined by \citet{Psaradaki:2023:AA} include troilite (FeS), pyrite (FeS$_2$), and magnetite (Fe$_3$O$_4$). They concluded that these compounds resulted in spectral fits that were less preferred. The oxidation state of Fe in troilite and pyrite is Fe$^{2+}$. For magnetite it is Fe$^{2+}$(Fe$^{3+}$)$_2$; hence, magnetite is likely to present XAFS that are quite different from the other compounds in our study. 
\citet{Westphal:2019:ApJ} analyzed observations from the sight line to Cygnus X-1 and worked to fit XAFS from FeS, metallic Fe, and silicate samples of suspected presolar dust grains. They concluded that less than $38.5\%$ of Fe resides in silicates; the rest is in metallic Fe with no signatures from FeS. 
If their finding that the Fe in the ISM is primarily in metallic form, then it would also dominate the observed $L$-shell extinction features and provide one of the better fits in our analysis.  We find that this is not the case, particularly for the Cygnus X-1 sight line that both we and \citet{Westphal:2019:ApJ} analyzed, which implies that metallic Fe is not the dominant species for this sight line.  This is in contradiction to their findings. Still, all X-ray absorption studies to date agree that Fe-S compounds do not account for an appreciable fraction of Fe-bearing dust grains. Our two best fitting compounds, lepidocrocite and hematite, are not present in prior works, warranting the future consideration of Fe$^{3+}$ oxides for the mineralogy of interstellar dust.

\subsection{Comparison to other X-ray models for interstellar extinction}
\label{subsec:ModelComparison}

In this work, we have presented a local {\tt XSPEC} model, \texttt{SolidFeL}, with dust extinction cross sections drawn from the laboratory work of \citet{Kortright:2000:PhRvB} and \citet{Lee:2009:ApJ}. The cross sections included in {\tt SolidFeL} are not shifted from the reported laboratory energies. The solid-phase, broad-band and photoelectric $K$-shell extinction features from elements other than Fe are modeled following the absorption profiles drawn from \citet{Hanke:2009:ApJ}, from which optical constants are then derived by \texttt{kkcalc} \citep[as described by][]{Watts:2014:OpticsExpress}. Below, we describe how {\tt SolidFeL} differs from other {\tt XSPEC} models of interstellar extinction.

The {\tt XSPEC} model \texttt{ISMabs} \citep{Gatuzz:2015:ApJ} provides high resolution theoretical cross sections for select interstellar metals. It models gas-phase photoabsorption only, and would be physically accurate if all interstellar metals were in the gas phase. The \texttt{tbabs} model also provides absorption-only properties of the ISM using the simplified cross sections for neutrals from \citet{Verner:1995:Ana}, but is modified for self-shielding by dust grains and follows the abundance tables of \citet{Wilms:2000:ApJ}. 
Both \texttt{ISMabs} and \texttt{tbabs} have shifted the metallic Fe cross sections of \citet{Kortright:2000:PhRvB} to match the observed position of interstellar Fe $L$-shell features \citep{Wilms:2000:ApJ, Gatuzz:2015:ApJ}. Both of these models are missing the dust scattering component to X-ray interstellar extinction. The updated version of {\tt ISMabs} \citep{Gatuzz:2015:ApJ} be found at \url{https://github.com/efraingatuzz/ISMabs}.

The {\tt XSPEC} model \texttt{xscat} \citep{Smith:2016:ApJ} solely models the scattering component of extinction from interstellar dust. This model contains the most comprehensive set of dust grain size distributions currently available in the X-ray, but does not account for dust grain absorption and shielding effects. The \texttt{ISMdust} model provides the full extinction profile -- absorption plus scattering -- from a power-law distribution of interstellar dust grains only, and is most accurate for telescopes that are able to resolve dust scattering halo images \citep{Corrales:2016:MNRAS}. It uses the empirical broad-band dust opacities of \citet{Draine:2003:ApJ}. \texttt{ISMdust} is best used in combination with an absorption-only model like \texttt{tbvarabs} or \texttt{ISMabs} to capture the gas-phase component of interstellar absorption, the latter of which was used in this work.

\section{Summary and Conclusions}
\label{sec:Conclusions}

The majority of interstellar Fe is injected in the gas phase by Type~Ia SNe, implying that Fe must condense into the solid phase in the diffuse environment of interstellar space \citep{Dwek:2016:ApJ}. This makes it difficult to draw conclusions about Fe-bearing interstellar dust from the mineralogical identifications of circumstellar dust. 
Additionally, there are no well-established solid interstellar Fe features in the IR. 
X-ray spectroscopy is thereby the only direct method for studying the mineralogy of interstellar iron.

We have examined archival {\it XMM-Newton} RGS spectra of Cygnus~X-1 and GX~339-4, which exhibit strong Fe $L$-shell extinction features around 0.7~keV. The spectra were fit with a non-thermal power-law model undergoing gas-phase absorption using \texttt{ISMabs} and dust-phase extinction using our \texttt{SolidFeL} model, which contains Fe $L$-shell extinction cross sections derived from \citet{Kortright:2000:PhRvB} and \citet{Lee:2009:ApJ}. We found that none of our available dust compounds provided a good fit to the spectra, unless we shifted the absolute energy scale of the laboratory data. We have determined the energy shift required for each species to best fit the observations. 

Of the five compounds tested, three provide equally good fits for different absolute energy shifts: the iron oxides lepidocrocite and hematite, and the iron end-member of olivine silicates (fayalite). 
Lepidocrocite and hematite are Fe$^{3+}$ minerals and have nearly identical XAFS cross sections that are unresolvable with modern day X-ray satellite spectrometers. 
Fayalite is an Fe$^{2+}$ mineral, and requires a larger energy calibration shift than the Fe$^{3+}$ oxides. 
We note that the energy shift required to fit the iron oxides to the data is within the standard deviation of the laboratory literature, whereas the shift required to fit fayalite requires a more significant departure from literature values. 
More importantly, a model in which fayalite accounts for the majority of Fe-bearing dust is physically inconsistent with the large abundance of Mg-rich silicates inferred from IR observations of astronomical sources and from in situ measurements of interstellar dust candidates captured in our Solar System. 
This makes Fe$^{3+}$ iron-oxide compounds our top choice as a major reservoir for interstellar Fe. Examination of other Fe$^{3+}$ minerals is warranted for future study.

The oxidation state of interstellar Fe is important, as it defines the starting conditions of Fe in protostellar clouds. If iron oxides truly represent the majority of Fe-bearing compounds in the ISM, that does not match the state of trace iron found in presolar and suspected interstellar grains collected in the Solar System. This conclusion would imply that interstellar dust must be significantly processed through the collapse of the molecular cloud that eventually forms a protostar and protoplanetary disk. To verify this hypothesis, careful calibration of absolute energy scale for metallic Fe and various oxidation states of solid Fe compounds is required, as that is the chief source of systematic uncertainty in this work.

Other conclusions from this study are summarized as follows.
\begin{itemize}
\item Including gas-phase Fe as one of our model parameters resulted in best-fit abundances that were consistent with zero.  These findings support that Fe is heavily depleted from the gas phase in the diffuse ISM.
\item The large parameter space of possible energy shifts and combinations of Fe-bearing dust species prevents us from drawing meaningful conclusions with multi-species fits at this time.
\item The total abundance of O and Fe seem over-abundant compared to solar values, while Ne appears slightly under-abundant. However, most results are within 3$\sigma$ of the uncertainties reported in \citet{Asplund:2021:AA}. 
\item The depletion and total dust mass fraction of carbonaceous grains is a strong source of systematic 
uncertainty for determining the total H column, which might explain some of the discrepancies in our abundance estimates.
\item From Fe $L$-shell absorption alone, we estimate that Fe-bearing materials account for $20-30\%$ of the total mass of interstellar dust. If this is the form of Fe oxides, then Mg-rich silicates would account for 13\% of the interstellar dust mass, leaving $\approx 60\%$ for carbonaceous grains. These conclusions are predicated on the hypothesis that there are distinct chemical populations of interstellar dust.
\end{itemize}

\acknowledgements

We thank the anonymous reviewer for their constructive comments. 
E.V.G., T.R.K., F.P., and D.W.S.\ were supported, in part, by the NASA Astrophysics Data Analysis Program (ADAP) grant 80NSSC19K0571. The work by L.R.C.\ and I.P.\ was supported through NASA ADAP grant 80NSSC20K0883. M.M.\ is grateful for funding of this project by the Federal Ministry of Education and Research (BMBF;  05K10GUB, 05K16GUC, and 05K19GUC) and by the German Research Foundation (DFG, project grant No.\ 510114039).  S.S.\ acknowledge financial support by the German Federal Ministry for Education and Research (BMBF, grant Nos.\ 05K19GU4 and  05K19RG3, respectively).

\facility{XMM}

\software{
SAS \citep{SAS},
XSPEC \citep{Arnaud:1996:ASPC}, 
ISMabs \citep{Gatuzz:2015:ApJ}, 
ISMdust \citep{Corrales:2016:MNRAS}, 
newdust \citep{newdust}, 
Astropy \citep{Astropy, Astropy2018, Astropy2013}
}

\bibliography{CygX1}{}
\bibliographystyle{aasjournal}

\end{document}